# ENHANCING SPACE SITUATIONAL AWARENESS TO MITIGATE RISK:

# A SINGLE-CASE STUDY IN THE MISIDENTIFICATION OF A RECENTLY-LAUNCHED STARLINK SATELLITE TRAIN AS UAP IN COMMERCIAL AVIATION

**Douglas J. Buettner**[(1)(2)(3)*], **Richard E. Griffiths**[(4)(5)], **Nick Snell**[(1)], and **John Stilley**[(1)]

[(1)]Department of Mechanical Engineering, University of Utah,
1495 E 100 S (1550 MEK), Salt Lake City, UT 84112, 801.581.6441, doug.buettner@utah.edu

[(2)]Acquisition Innovation Research Center, Stevens Institute of Technology, 1 Castle Point Terrace, Hoboken, NJ 07030, USA, 201.216.8300, airc@acqirc.org

[(3)]Sr. Member of the AIAA

[(4)]Department of Physics & Astronomy, University of Hawaii at Hilo, 200 W. Kawili St., Hilo, HI 96720, USA, 808.339.4760, griff2@hawaii.edu

[(5)]Department of Physics, Carnegie Mellon University, 5000 Forbes Ave., Pittsburgh, PA 15213, USA, 808.339.4760, rgriff@andrew.cmu.edu

∗To whom correspondence should be addressed

## ABSTRACT

Over the past several years, the misidentification of some SpaceX/Starlink satellites as Unidentified Aerospace Phenomena (UAP) by both pilots and laypersons has generated unnecessary aviation risk in the air and confusion on the ground. The many deployment and orbital evolution strategies, coupled with ever changing sun specular reflection angles, contribute to this gap in space situational awareness. While SpaceX/Starlink and other satellite operators are working on partial mitigations of this novel light pollution for the ground-based astronomical community, it is unlikely that these mitigations will resolve the misidentification of all Starlink configurations viewed from the air, nor those viewed from the ground by the general public. In this paper we present an astrometric analysis of an incident that generated multiple, corroborating reports of "a UAP" from five pilots on two commercial airline flights over the Pacific Ocean on August 10th, 2022[1]. This incident included two cell phone photos and a video of an unrecognized and possibly anomalous phenomenon. We then use supplemental two-line elements (TLEs) for the Starlink 'train' of satellites launched that same day and Automatic Dependent Surveillance-Broadcast (ADS-B) data from the flight with the photographs to reconstruct a view of these satellites from the aircraft cockpit at the time and place of the sighting. The success of this work demonstrates an approach that could, in principle, warn aviators and the public about satellites that could be visible in unusual or novel illumination configurations, thus increasing space situational awareness and supporting aviation safety. The approach is based on standard orbital mechanics and ray-traced rendering of the view from the pilots' cockpit in visualization simulations. In the implementation of our approach, we were able to closely match the apparent speed in degrees per second traveled by the object between the two photographs taken by one of the pilots. Further, our rendering experiments suggest that the Starlink satellite train would not have been visible if the solar arrays were not deployed. We then discuss the potential for this visualization approach to reduce unknown sightings from Starlink satellites or other man-made space objects that can lead to pilot distraction and subsequent airwave chatter. We describe future work to fully simulate the range of deployment configurations with the sun's relative location effect on the illumination of and reflections from these objects. We conclude with recommendations for government and satellite operators to provide better *a priori* data that can be used to create advisories to aviators and the public. The automated simulation of the reflection of light off satellites could also support researchers investigating sightings of unfamiliar aerial/aerospace objects as likely being from normal versus novel space events.

---

[1] One of the flights had a pilot who was in training, hence the total pilots observing the object were the captain, the co-pilot, and the pilot in training.

## 1. Introduction

Space Exploration Technologies Corporation (SpaceX) Starlink satellites, often misidentified as Unidentified Aerospace Phenomena (UAP) [2], have generated significant press over the past few years (CBS Pittsburgh 2021, WCVB Channel 5 Boston 2023, Reyes 2023, Tangermann 2019, Grassi 2023, Mandelbaum 2019). While SpaceX is working to at least partially resolve this novel light pollution issue for the astronomical community (Tangermann 2019, Loeffler 2023, SpaceX 2022), we do not believe it will resolve the misidentification as UAP issue as recent videos of newer Starlink "2.0 minis" are still clearly visible (Schrader 2023, LIVE 2023).

In addition, recent US government congressional hearings about UAP and concomitant media attention about the alleged recovery of craft of extraterrestrial origin (117th Congress (2021-2022) 2022, C-SPAN 2023) has intrigued the public. The U.S. military, under congressional order, has set up a special office for collecting and addressing UAP reports from all current or former government employees, service members, or contractors (AARO 2024). In early 2024, a congressional bill related to flight safety was introduced that will help facilitate reporting of UAP sightings to the Federal Aviation Administration (FAA) by all commercial aviation personnel (Garcia and Grothman 2024, Thomas 2024). As a result of these activities, people from many sectors are becoming acutely aware of what they are seeing in the sky, such that if they do not know what it is, they are now providing cell phone photographic and/or video evidence of their sighting. This increased awareness of an unresolved evidence trail of UAP sightings, and the consequent risks to aviation safety and national security, has resulted in a concerted effort to look beyond the dismissal of sightings as simply social and psychological phenomena (Sharps 2023, Torre 2024, Yingling 2023a, Yingling 2023b).

This is leading to a reduction of the stigma associated with the serious study of phenomena underlying 'Unidentified Flying Objects (UFO)' (Sultan 2023, Dunn 2023, Gallaudet and Mellon 2023, Williamson 2023). The once reluctant scientific community is slowly starting to consider the underlying causes of UAP reports as an interesting area for research (Yingling 2023a, Yingling 2023b, David 2023, Medina 2023, Watters, et al. 2023), with global reach (Lomas 2023). However, we lack sufficiently mature tools and methods to systematically analyze sighting reports for even the "known phenomena" category. Without such critical methodologies in place, satellites, rockets, and drones will continue to clutter pilot communication channels and aviation hazard databases as these knowable lights in the sky are reported as UAP.

---

[2] We adopt the use of Aerospace in the UAP acronym for this paper vs the use of Aerial or Anomalous to align with the Scientific Coalition for UAP Studies' (SCU) use (Scientific Coalition for UAP Studies 2021).

The approach described in this paper could help remove a significant fraction of known space-related events from being mischaracterized and misreported as UAP. UAP reports related to Starlink satellites collected by citizen UAP reporting centers in 2020 comprised 18% (MUFON) to 25% (NUFORC) of all reports (Cockrell, Murphy and Rodeghier 2023). In addition, the results of a Freedom of Information Act (FOIA) request to the Federal Aviation Administration (FAA)[3] resulted in a list of 69 reports identified as "UFO-UAP" from January 23rd, 2023 to April 27th, 2023. From this list, an initial estimate is that 36% or more of these could be Starlinks; a more detailed analysis using this approach may be attempted in future work.

We start this paper with a case study of reports of an unidentified aerial object observed over the Pacific Ocean by five airline pilots (Pittet, 2023), all of whom are considered "trained observers" (Graves 2023a, Schwartz 2022). One of these pilots also provided photographic evidence of the UAP, which appears as a long, slender, and solid white object. They did not consider that these were a Starlink satellite launch, which are typically believed to manifest as a "string of pearls" (Lalbakhsh, et al. 2022), as from their perspective they were viewing the Starlinks at an angle of about 55 degrees between their relative velocity vectors. Here we provide the raw observations that were reported to the worldwide Mutual UFO Network (MUFON; case number 124190) and the subsequent analysis using astrometry to characterize this UAP case using the available evidence at that time. We include this exercise as a case study for extracting information about anomalous objects from photographic and astrometric evidence as an important step in support of subsequent research using modeling and therefore, the advancement of the scientific study of UAP.

Following the photometry study, we describe a method that uses tools from orbital mechanics and visualization simulation techniques that can allow researchers to filter out low-altitude satellites such as Starlinks and other man-made space objects from UAP reports. This enables the removal of some fraction of identifiable objects from the larger "unknown" category of reports. As this method is in its infancy, we need to make assumptions about these satellites (e.g., spacecraft attitude which is the 3-dimensional orientation in space, the surface material's reflective properties and the deployment of solar arrays, antennas or other mechanical deployables). We conclude with recommendations regarding additional data required to support building an analytic capability that could provide notices in advance to aviators and the public about known man-made space-related events that would reduce the number of known objects reported as UAP[4]. We also identify future research directions that would support both *a priori* notifications and after-the-fact analysis of UAP sightings and reports.

---

[3] Request was filed by Robert Powell as documented here (Powell 2024) while the response letter and the results have been uploaded to our GitHub site (DrDougB 2024).

[4] We wish to further highlight the "known man-made" aspect of our study. There are reports from pilots, for example (Graves 2023a, Graves 2023b) and ground observers (MUFON 1969) that we believe cannot likely be explained using the orbital analysis methods presented here. However, we feel the general physics-based 3D modeling of events may in fact support eventual characterization of some various classes of UAPs; no matter what the true nature of their origin is.

## 2. The Observational Case Study

**Observations**

Here we document the observations that were made from two separate aircraft flying on the same flight path over the Pacific Ocean in August of 2022. The lead aircraft was flight number AC536 flying from Maui to Vancouver B.C. The trailing aircraft was flight number AC34 flying from Sydney Australia to Vancouver B.C.

Four or five very bright unresolved star-like objects were seen by the captain and co-pilot of AC34, as reported by the airline captain (Pittet 2023). The individual objects brightened to be brighter than the International Space Station for a short period of time and then faded. The captain provided an estimated distance of 35-75 nautical miles when first sighted on his assumption that they were within the atmosphere at an altitude which was roughly the same as the airliner (a barometric altitude of ~39,000 ft). The captain watched these individual star-like objects fade and appear to converge into a single large, craft-like object.

This craft-like object then appeared to fly ahead and roughly parallel with the aircraft's flight path, at which point it came into the field of view of the leading aircraft AC534, on the same flight path. From this lead aircraft, the large craft could be observed moving parallel and just above the horizon, apparently at 37,000 ft., traversing an azimuthal angle of 90 degrees (from North-West to North-East) in about two minutes.

During the latter part of this flight, the apparent large craft ('cigar shaped' in projection) was photographed and videoed by an anonymous pilot of the lead plane (flight AC536) and confirmed visually by two other pilots on this same flight. The initial bright objects were not seen by the pilots of this latter flight because the objects were directly behind them, and the two aircraft were separated by about 160 miles.

The integrated brightness of the elongated object in the photos/video from the lead plane was very much less than the apparent integrated brightness of the 4 to 5 objects originally observed by the trailing plane.[5]

In all, observations of the large object were confirmed by five pilots with two cellphone photos (Figure 1 shows photo 1 taken at 2022-08-10T11:39:08UTC and Figure 2 shows photo 2 taken at 2022-08-10T11:39:24UTC) and a ~16-second video taken by the pilot on the leading plane. The video was bracketed by the two photos. The initial, separate, star-like objects were seen by the two pilots of the trailing plane only. The approximate longitude/latitude of the aircraft at the time of sighting was 39 degrees North, -133 degrees West, as determined from 'FlightRadar24'[6]. There were no radar records of any of the objects that were seen visibly by the pilots – the airliners (a Boeing 789 and a Boeing 38M) had onboard radar tuned for weather (forward-looking), but not for objects at large azimuthal angles relative to their heading vectors.

---

[5] This is based on the description of the objects as originally witnessed by one of the pilots.
[6] Found online at: https://www.flightradar24.com/

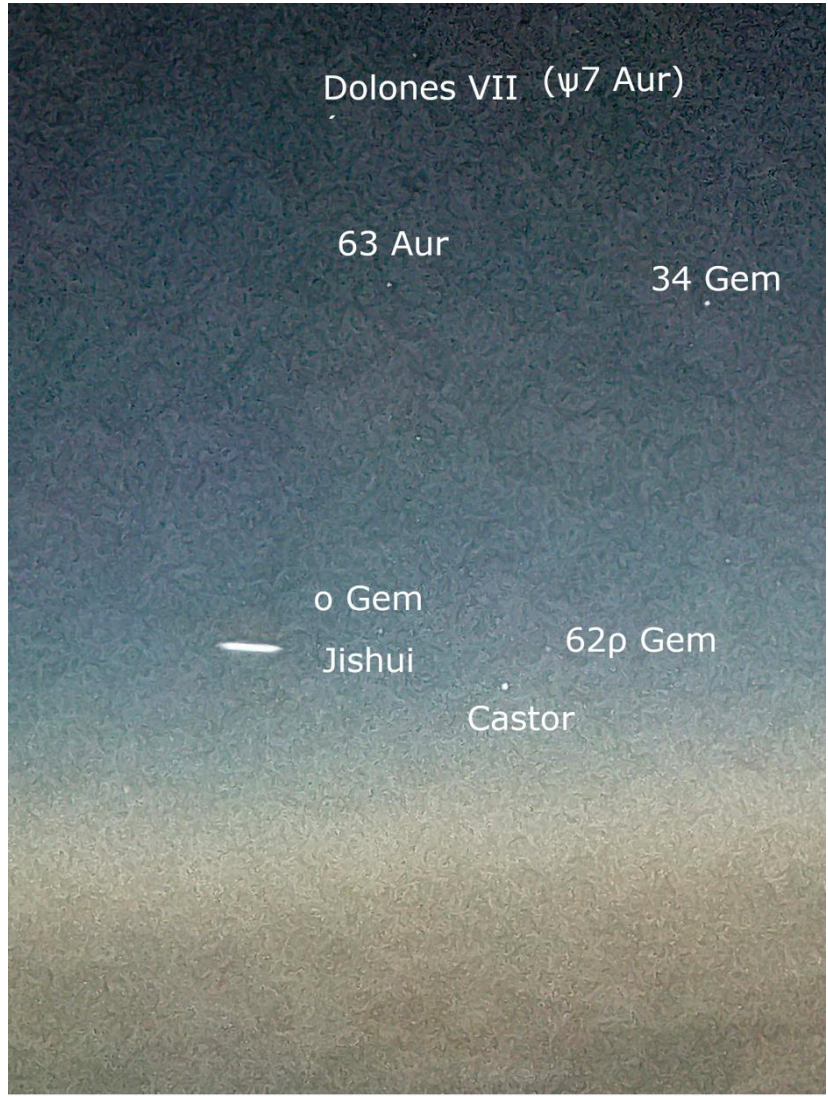

**Figure 1. Photo 1 from an iPhone 12 camera taken at 2022-08-10T11:39:08UTC as determined by information embedded in the JPEG[7] metadata. The astrometric solution to the starfield, i.e. the identification of the stars, was found using the publicly-available software package Astrometry.net. The object was observed at ~2 am local solar time when the aircraft was at a latitude of 39.60321 degrees North, and a longitude of 138.436 degrees West.[8]**

---

[7] JPEG (Joint Photographic Experts Group) is a common image compression algorithm used on most cell phones (Wikipedia 2024a).
[8] The aircraft's latitude and longitude are extracted from the Satellite Orbital Analysis Program (SOAP) in this figure and Figure 2. SOAP is discussed later in this paper.

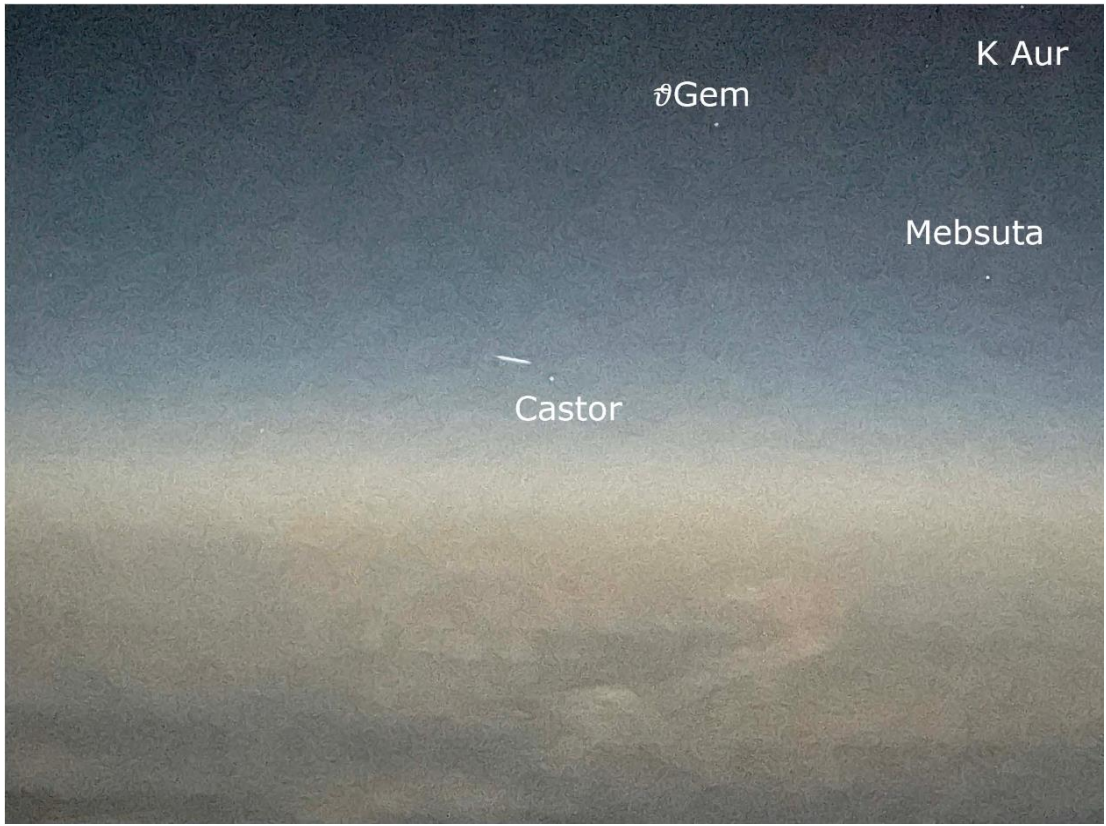

**Figure 2. Photo 2 from the iPhone 12 camera taken at 2022-08-10T11:39:24UTC as determined by information embedded in the JPEG metadata. An astrometric solution was unable to be provided by Astrometry.net as it failed to identify the stars in the image due to the camera's zoom factor distortion. The astrometric solution was provided manually in the second image based on the first image's solution. At this time, the aircraft was at a latitude of 39.62299 degrees North, and a longitude of 138.404 degrees West.**

The cellphone video lasted nearly 16 seconds, but at 12.7 seconds the cellphone was rotated through approximately 90 degrees because of aircraft cockpit window constraints. The recorded intensity of the object then faded for about 0.5 seconds during the cellphone movement and rotation. The image of the object was recovered but the focus was not recovered by the end of the video at 16 seconds. The useful part of the video therefore lasted for the first 12.7 seconds. No stars were recorded in the video because of the short exposure times per frame.

**Photometry**

The initial sighting was of 4 or 5 very luminous point source objects seen close to the NW horizon with individual objects estimated to have a brightnesses of about 2 or 3x brighter than the International Space Station (ISS) or Venus, with a magnitude of approximately -5. At the estimated distance of 35-75 nautical miles, and assuming isotropic emission, the luminosity of each object was at least 300 kW. There is no photographic evidence for this initial sighting. Analysis of the two photos taken from the leading plane shows independent confirmation of the brightness of the large object, but not the original star-like luminous objects.

The two cellphone JPEG images (from an iPhone 12), shown in Figure 1 and Figure 2, were converted to .fits (Flexible Image Transport System) files for examination using

SAOimage DS9[9] from the Smithsonian Astrophysical Observatory, so that the RGB images could be examined separately. SAOImage DS9 (Mandel 2003) is widely used by astrophysicists for data analysis of astronomical objects against a dark sky background and is an appropriate software package to use in this case. Other image analysis software packages such as 'Forensically' are commonly used for everyday cellphone photography in daylight (for verification purposes that the image is a 'bona fide' image and not a 'fake') but are not as useful in this case, where there is a bright object against a dark sky background, as commonly occurs in observational astronomy. Photometric analysis can be performed with DS9.

To get an approximate value of the brightness of the object, we can examine the photo images and see that the digital pixel values (analog-to-digital units, ADU) are about 250 per pixel within the image of the large object, which is about (115-120) x 13 pixels in size as shown in Figure 1, with a background intensity of 140 per pixel, i.e. the integrated intensity value is about 166,000 ADU.

This compares with an integrated pixel intensity value of 14,000 ADU for the nearby star Castor, which has (Pogson) magnitude (mag) 1.58. Hence, the integrated mag of the large object (11.5 times brighter than Castor) is about 5.5 mags brighter, i.e. a mag of -4.

The large object, as shown in Figure 1, appears to be "cigar shaped", with scalloped protrusions at either end, with an overall aspect ratio of approximately 116:13. At the time the cellphone images were taken, the distance to the object was not known, but the pilots estimated it to be about 20-30 nautical miles from the plane at roughly the same altitude and within the atmosphere (in this case at a barometric altitude of ~37,000 ft).

The delay between the first sighting from the trailing plane and the first photo from the leading plane is not known but was more than a minute: the captain of the trailing aircraft observed the initial objects in the NW, while the video and photos were taken from the leading plane towards the NNE, as shown by the identification of stars in the background (see discussion of astrometry, above). An approximately 1.5-degree apparent angular size, Figure 3, corresponds to a linear size of about a mile for 30 nautical miles in distance.

---

[9] Found online at: https://sites.google.com/cfa.harvard.edu/saoimageds9

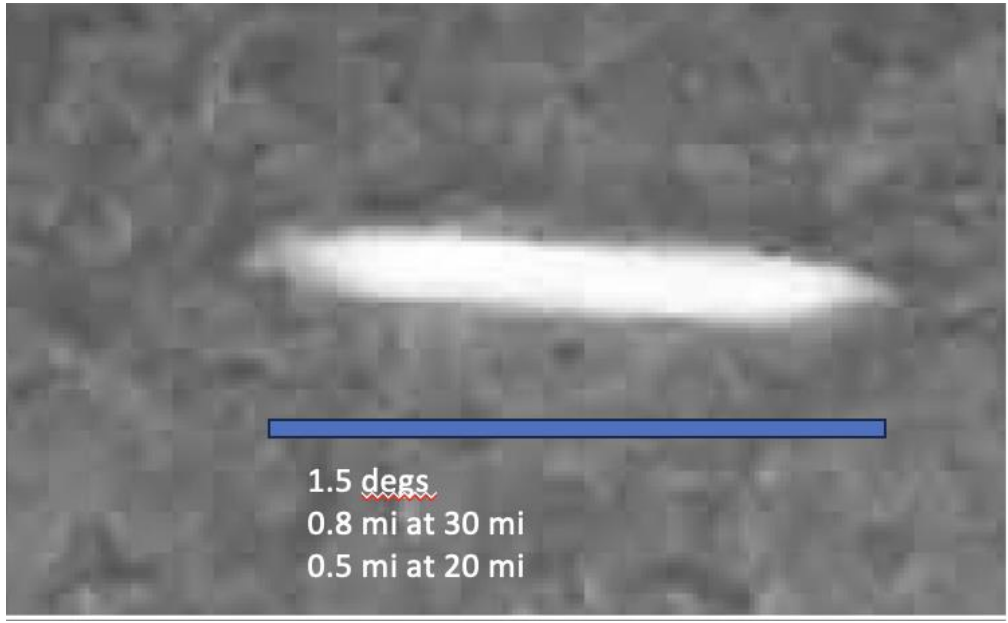

**Figure 3. Using the astrometric solution for the background stars, we estimate the apparent angular size of the object as 1.5 degrees in Photo 1. The apparent linear lengths are based on estimated distances of 30 and 20 nautical miles, resulting in linear dimensions of 0.8 and 0.5 miles respectively.**

**Angular Resolution and Apparent Sizes**

The images in Figure 1 and Figure 2 were taken with an Apple iPhone-12. The iPhone-12 uses lenses of about 2 mm diameter, and therefore has a diffraction limit of about 2 arcminutes (about 2 pixels). No smaller structures can possibly be resolved. The overall size of the cellphone images is approximately 1530 x 2040 pixels, covering 24 x 32 degrees, so that the pixel size (plate scale) is approximately 1 arcminute, smaller than the diffraction limit.

The object appeared in the sky as cigar-shaped, about 116 pixels long (1.5 degrees) and 13 pixels wide in Figure 1, compared with about 5 pixels width for a stellar image. Therefore, the object is only partially resolved in the short (vertical) dimension. In the long dimension, no substructure is visible in the images at the cellphone optical resolution, except for the two ends of the 'cigar', which both seem to be scalloped on their lower sides (like an aircraft carrier). Alternatively, the object may have apparent extensions which are narrow in height. Photo 1 and the video of the object do, however, show structural subcomponents.

The object was observed at about 2 a.m. local solar time. Assuming that the object was within the atmosphere, at the estimated distance of 20-30 nautical miles away from the aircraft from which the pictures were taken, the luminosity was therefore presumed to be intrinsic to the object. From the perspective of the aircraft the sun was about 30 degrees below the horizon, and the moon had recently set.

In Figure 2, the longitudinal axis of the object 'disk' was not parallel to the horizon but had an apparent negative pitch angle of about 5 degrees, i.e., leading edge pointing downwards. There was no other indication that the object was descending. Neither is there any indication from the photos or video of any aerodynamic features (e.g. wings, or tail). The downward pitch of about 5 degrees can lead to imaginary features in the low-resolution image, see for example pixel jumps on the topside of the object, but such features are simply caused by 'pixelization' of the image in the cellphone sensor and are not real.

Upon examination of the separated RGB images, SAOImage DS9 shows that there was no change in color across either the narrow dimension or the long dimension – i.e., there was no indication of heat generation anywhere along the length of the object. Some frames of the video do, however, show a gap along the length of the object (see Figure 4 as an example of the observed gap structure).

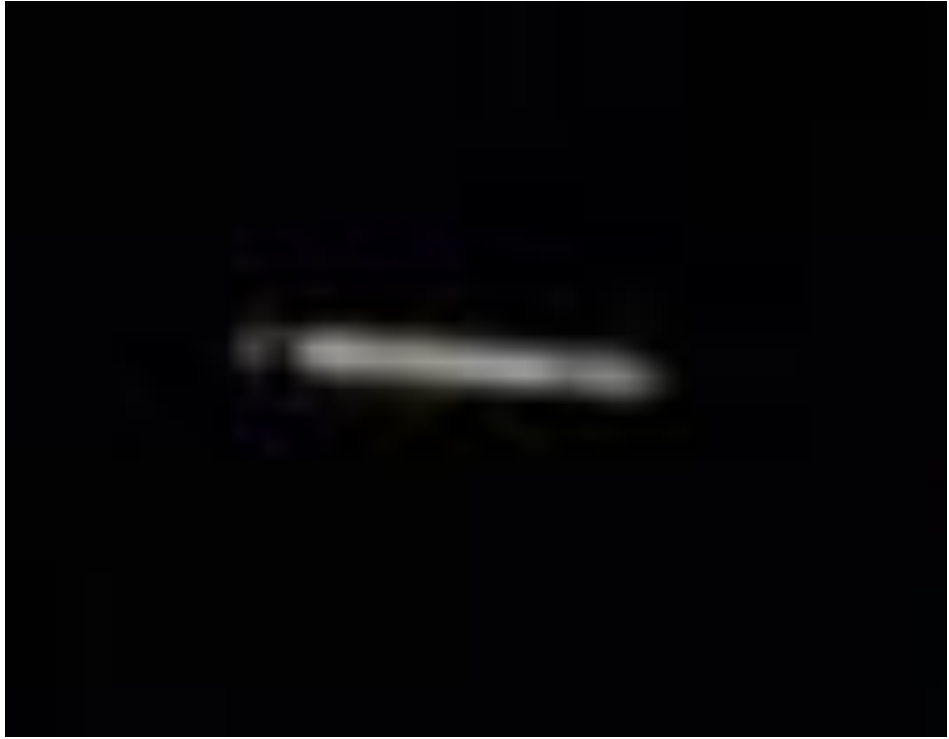

**Figure 4. Frame # 26-04 from the pilot video, at about 8 seconds after the first photo, showing a gap structure**

The atmospheric (wavelength) dispersion of the image perpendicular to the horizon should amount to several arc seconds between blue and red images, but this is not apparent either in the stellar images or that of the observed object: the cellphone image pixel size is one arc minute, so the atmospheric dispersion is not resolved. Refraction by the atmosphere should, however, be significant, such that stars near the horizon appear to have higher altitude than they would have with no atmosphere (e.g. the actual sun at sunrise or sunset is depressed about half a degree relative to the apparent sun).

**Celestial Coordinates and Apparent Velocity**

The object was first observed from the trailing aircraft above the North-West horizon, and moved parallel along the horizon towards the North-East, roughly following the flight paths of the two Air Canada aircraft. The object crossed the flight path vector of the leading plane but was by then beyond the view of the trailing plane. The cellphone photos and video were taken from the (leading) flight when the object was above the North-North-East horizon and projected against the constellation of Gemini. There are stars visible in both photo images, though not in the individual video frames because of the short effective exposure times. Astrometry on the stars in the images was performed using astrometry.net (Dustin Lang et al. 2010).

The following analysis is from the lead aircraft photos and witness report: Astrometry of the stars in Figure 1 shows that the object, as observed from the leading aircraft, was projected on the sky at an apparent position of Right Ascension (RA) 7h 51m, declination (dec) +36d,[10] about 2.4 degrees west and 0.5 degrees in elevation south of Jishui, 71-$o$ Gem. The video shows that the apparent size of the object did not change during the 12.7 second useful interval of the video. Photo 2, however, shows the object angular size to be smaller than in Photo 1, because a 'zoom' factor has been applied to the second cellphone picture. This zoom factor of about two means that the overall image size of Photo 2 is too wide for application of the astrometric program Astrometry.net, which 'typically' fails with such wide-angle cellphone images because the program assumes a tangent-plane projection for the images and does not take into account spherical projection effects (Lang 2023), or distortion in the wide-angle cellphone images (i.e. pincushion and/or barrel distortion), visible by inspection of the shape of the horizon.

Manual astrometry of the stars in Figure 2 showed that the object had moved to an apparent position of RA 7h 36m, dec +32d, about 0.5 degrees in elevation above and 1.2 degrees west of Castor. Other stars recorded in Figure 2 are identified as $\tau$-Gem, and $\kappa$-Aur. We include an image for the constellation Gemini in Figure 5 for reference. The camera image axes are taken to be roughly altitude-azimuth as shown.

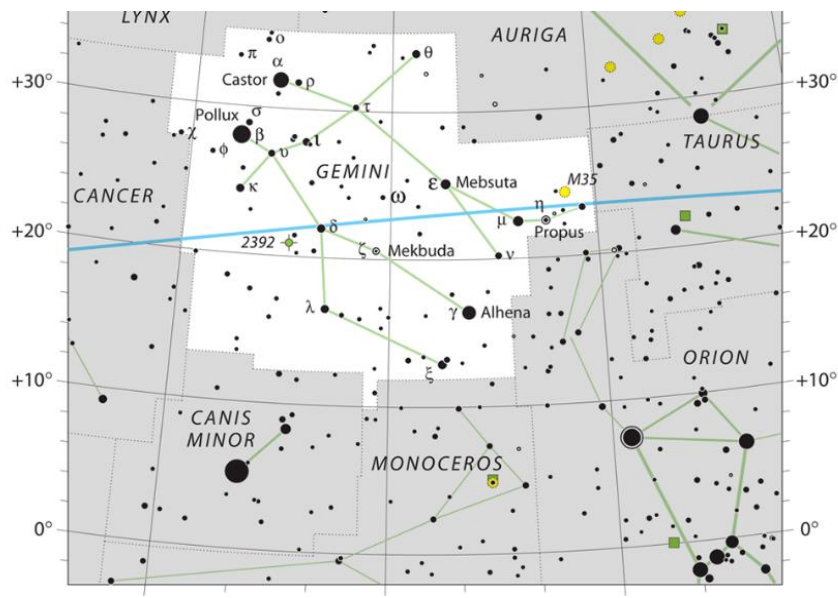

**Figure 5. The Gemini constellation (IAU and Sky & Telescope 2015).**

The angular distance traversed by the object between the two photos was therefore approximately 4.22 degrees, and the time difference was 16 seconds, as determined from the JPEG files' metadata. Neglecting the motion of the aircraft, the angular rate of motion was thus 0.26 degrees per second, or 90 degrees in just over 2 minutes, in agreement with the witness statement that the object moved 90 degrees across the sky (NW to NNE) in about 2 minutes. The object, as observed from the leading aircraft, had an estimated distance of about 30 nautical miles (somewhat less than the 35-to-75-nautical mile distance of the initial bright lights seen from the trailing aircraft). At this assumed distance, the angular rate of motion then translated to about 500 mph.

---

[10] This is standard astronomical notation for hours (h), minutes (m) and degrees (d) for celestial coordinates.

Allowing for the motion of the witness aircraft, the apparent tangential speed of the UAP was thus about 1100 mph, with no measurable radial velocity towards or away from the aircraft.

## 3. Modeling Starlink Group 4-26

### Orbit Simulation and Visualization

We then investigated whether this sighting could have resulted from a launch of SpaceX/Starlink satellites made earlier the same day (August 10, 2022 at 02:14 UTC) from Pad LC-39A from Kennedy Space Center in Florida USA, i.e. Starlink-54 Group 4-26 (Wall 2022, Sesnic 2022, RocketLaunch.Live 2022). To perform orbital modeling of this satellite group, we use Two-Line Elements (TLEs) for each of the 52 satellites.

Two-Line Elements (TLEs) are a common NORAD format and incorporate the Keplerian elements with effective time and date information that describe the orbital variables for the satellites and rocket debris (Kelso 2022d, Wikipedia 2023a). We first used the Satellite Orbital Analysis Program (SOAP) version 15.4.1 from The Aerospace Corporation (Aerospace), (Aerospace 2020). However, as SOAP is only available to employees of Aerospace and their government customers, we also used a commercially available program, System Tool Kit (STK) version 12.6.0 (AGI/ANSYS 2024) which can perform similar orbit determination analysis as SOAP.[11]

We also investigated tools capable of providing a visual representation of what the pilots in the aircraft from which the photos were taken would have seen. For this purpose, one of us investigated several tools, settling on the open-source tool Blender 4.0 (Blender 2024) for this purpose due to its price (free), capabilities, and support community even though it is primarily used by the animation community. Blender allowed us to load a CAD model of the aircraft (Goo 2022), model the earth[12], and represent the sun's lighting with the ability to perform ray tracing (blender 2023).[13]

### Obtaining TLE, ADS-B and Other Data

TLE's compatible with SOAP were pulled from Celestrak (Kelso 2022a) which includes the names of the individual Starlink satellites. Supplemental TLE data for the objects in this Starlink group were also provided by Jonathan McDowell of the Smithsonian Astrophysical Observatory (private communication). Subsequently, Kelso updated CelesTrak's supplemental query of Celestrak's NORAD archives for "after the fact queries" to extract a Starlink satellite group for a short time frame (7-10 days) using a 7XXXX query format where the XXXX represents the Starlink satellite number (see Figure 6). Hence, a Starlink satellite named 'STARLINK-1234' would be 71234 in Celestrak's supplemental query (Kelso 2022c). The combined TLE file used in SOAP is included on our GitHub site (see section 9. Supplementary Materials provided later in this paper). The Starlink satellite name is obtained from Celestrak's satellite catalog by searching through the catalog for the satellite's associated with the UTC launch date (Kelso 2024a). A screenshot of the query page containing the inputs for Starlink group 4-26 is provided below in Figure 6. At the time of this article, this is the form

---

[11] While there are several capable orbit simulation tools available, we chose to use STK due to its position in the marketplace and use by aerospace engineering companies.
[12] Uses a spherical earth and imagery from NASA (NASA 2004).
[13] We elected to forego trying to use the Blender plugin to use the sun for the light source as we do not know the materials used; electing to simply model the satellites as 3D rectangular shapes with a surface that emits light. We describe this rendering limitation in later sections of this paper.

individuals should use with their name and e-mail address to perform a similar query (Kelso 2024b).[14,15]

**Figure 6. Screengrab of Celestrak's NORAD archive Supplemental Query Request Form (last accessed in March 2024)**

In an *a priori* pull, prior to a launch, using for example our Starlink Group 4-26 launch, the user would instead use a Celestrak query formatted as follows:

https://celestrak.org/NORAD/elements/supplemental/sup-gp.php?FILE=starlink-g4-26&FORMAT=TLE

This will return the Starlink payload as "STACK" and the second stage as "SINGLE" using 72000 and 72001 identifiers in the TLE information prior to NORAD ID assignment. This query result is:[16]

```
STARLINK-G4-26 STACK
1 72000C 22097A   22222.10427778  .00079168  00000+0  13040-3 0    09
2 72000  53.2190 249.4977 0077526  44.7868  32.1202 15.96675264    19
STARLINK-G4-26 SINGLE
1 72001C 22097B   22222.10427778  .01041707  00000+0  16879-2 0    01
2 72001  53.2189 249.4977 0077464  44.9070  32.0000 15.96664287    14
```

---

[14] Dr. Kelso maintains a Twitter account for announcements regarding changes to Celestrak. He recommends that individuals should monitor this account for changes to Celestrak's website.
[15] In some recent cases, we have also used SAIC's website (SAIC 2019) to pull data for objects. However, we found Celestrak's founder to be extremely responsive to our requests for additional query functionality in support of this effort.
[16] To understand the TLE format, please refer to references (Kelso 2022d, Wikipedia 2023a).

After a launch has occurred, one needs to pay attention for updates to the Celestrak satellite catalog to identify the correct NORAD IDs for the objects.

It is important to note the following: "The best we do, by working with SpaceX, is to propagate from the deployment time (when thrusting has stopped) and then fitting that propagation with SGP4[17] to produce pre-launch 'SupGP' data. Within 8-16 hours we usually have individual ephemerides for each satellite that then supplants the pre-launch estimate." (Kelso 2023, Hoots 1988, Vallado, et al. 2012, Kelso 2022b). This forces us to manually select the closest UTC time to the photographs for each individual satellite from the query results.[18] Recently, Kelso has made further updates to better support providing supplemental data when queries are being made close to but after a launch has been made (Kelso 2024).

To determine the location of the aircraft during their flights, we used the Automatic Dependent Surveillance-Broadcast (ADS-B) data from both flights discussed in the case study. However, for our orbit determination and visualization analysis in SOAP, we were more concerned with aligning our simulation to the flight with photographic evidence (AC356).[19] Hence, we only used the ADS-B data from the leading flight, Air Canada (AC536) flight from Kahului, Maui to Vancouver, Canada in our simulations. In addition, since the UTC (Coordinated Universal Time) time for the photographs were available from the photograph's meta-data, we only included the ADS-B for the portion of that flight that was critical for our modeling effort. Hence, the ADS-B data used in SOAP is from 2022-08-10T10:44:22UTC to 2022-08-10T12:49:30UTC.[20]

The ADS-B data for this flight has barometric pressure-derived altitudes in feet. SOAP, however, requires altitudes above mean sea level (MSL) in kilometers. To accommodate the difference between barometric altitudes and MSL we converted the barometric height values to the Geoid using Orthometric heights from an online conversion tool provided by the National Science Foundation's Geodetic Facility for the Advancement of Geoscience (GAGE) (EarthScope Consortium 2023).

Finally, the modeling efforts for STK and Blender incorporated a Computer Aided Design (CAD) model of a Boeing aircraft to simulate the appearance of the satellite train more accurately from the aircraft's cockpit. In addition, STK incorporated a CAD model of the SpaceX Falcon 9, while Blender also incorporated a GeoTIFF image of the Earth. The online repository of our modeling results provides these models (see *section 9. Supplementary Materials*).

**Orbit Simulation Results**

Using the times that the two photographs were taken we provide the following graphics from SOAP in Figures 7 to Figure 11.

---

[17] SGP is an acronym for Simplified General Perturbations (SGP) where SGP4 is one of the models used by orbit propagation software such as SOAP.
[18] We anticipate that future process improvements could automate most of this currently manual process.
[19] ADS-B data for the flights was provided by Dr. Sarah Little.
[20] The file "AC536.xlsx" on our GitHub website has the original ADS-B data for the flight and includes all associated numerical calculations used to incorporate AC536 into SOAP's orbit simulation of Group 4-26.

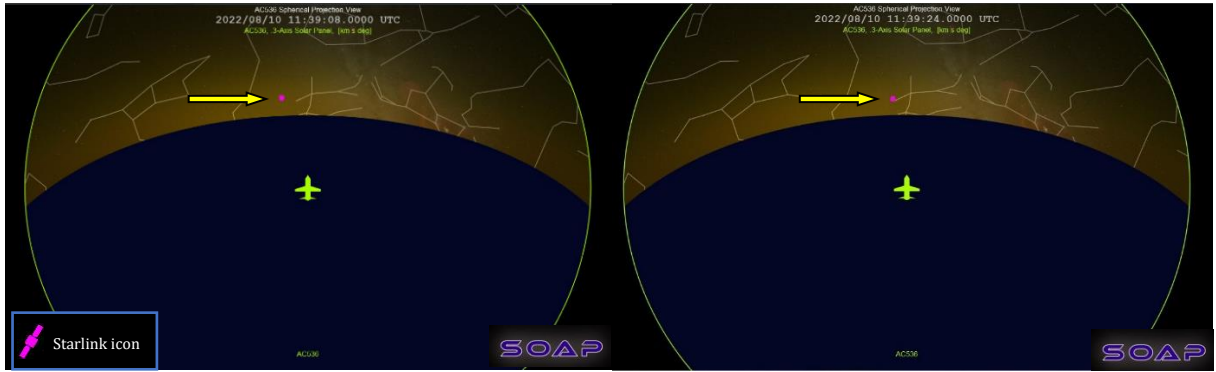

**Figures 7a and 7b. Screenshots of SOAP's Spherical Projection View[21] viewed looking down on flight AC536 (using SOAP's yellow aircraft icon with the size exaggerated to make it clearly visible) from above to clearly show the constellations visible to the aircraft above their horizon with respect to the aircraft heading. The satellites (the purple satellite icon) are the positions of all 52 satellites that have been propagated to their orbital location at the same UTC time of the two photographs, yellow arrows are included to help identify their location in the screenshots. Fig 7a, the image on the left has the satellites propagated to the UTC time for the first photograph (Aug 10th, 2022 11:39:08UTC), while Fig 7b, the image on the right has the satellites propagated to the time for the second photograph (Aug 10th, 2022 11:39:24UTC). The satellites are clearly near Gemini.**

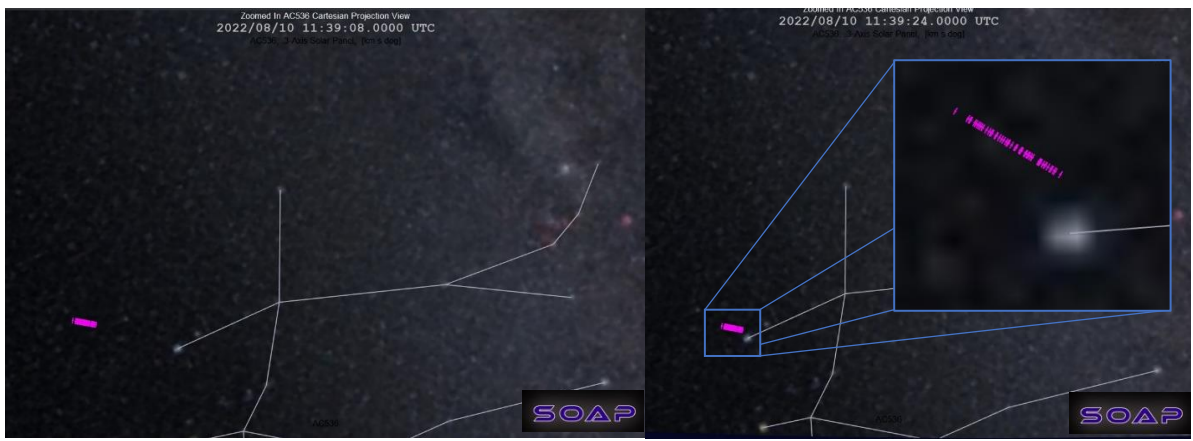

**Figures 8a and 8b. Screenshots of SOAP's Cartesian Projection View of the zoomed in relative locations of the satellites (same purple satellite icon as used in Figures 7) as viewed from the cockpit of AC536 aircraft at the time of the photographs, again clearly showing its location with respect to Gemini. Fig 8a is from the UTC time for the first photograph (Aug 10th, 2022 11:39:08UTC), while Fig 8b has the satellites propagated to the time for the second photograph (Aug 10th, 2022 11:39:24UTC). We also include in 8b a further zoomed in insert to show the gaps in between the satellites as viewed from flight AC536's cockpit.**

---

[21] If AC536 were above land instead of the open ocean, this view in SOAP would have shown the land features on a spherical Earth.

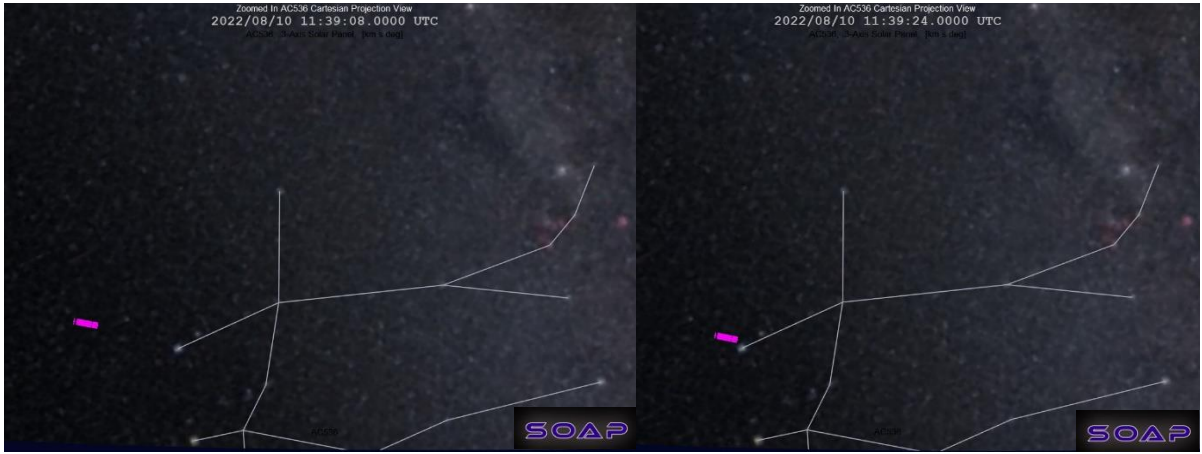

**Figures 9a and 9b.** This figure is a repeat of Figures 8a and 8b which are SOAP's Cartesian Projection View of the zoomed in relative locations of the satellites (same purple satellite icons). However, in these figures the aircraft's position has an additional 0.8 km included in the altitude to account for potential barometric altitude error. The zoomed in insert in Figures 8b has not been included in Fig 9b. These screenshots are provided to demonstrate the imperceptible visual effect of an altitude error in our results.

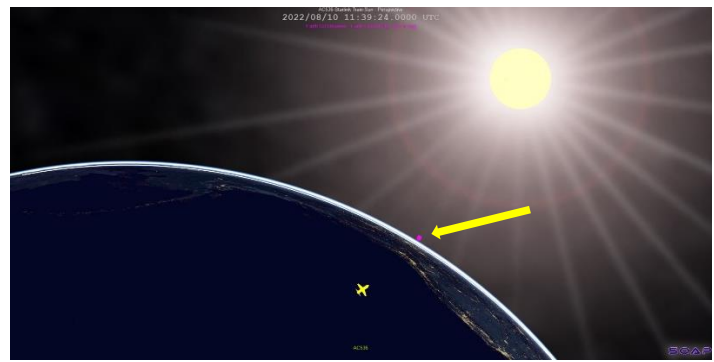

**Figure 10.** Screenshot from SOAP'S spherical view with the Starlink satellite train propagated to the time of the second photograph showing it in full view of the sun. This viewpoint is representative of an observer at a much higher altitude and is used to show that the satellite train is in full view of the sun. The yellow arrow is included to aid finding the location of the satellite train in this image. The yellow aircraft icon for AC536 shows the direction the aircraft was traveling and that it is still in the Earth's shadow, where the white line along the Earth's limb is the day/night terminator.

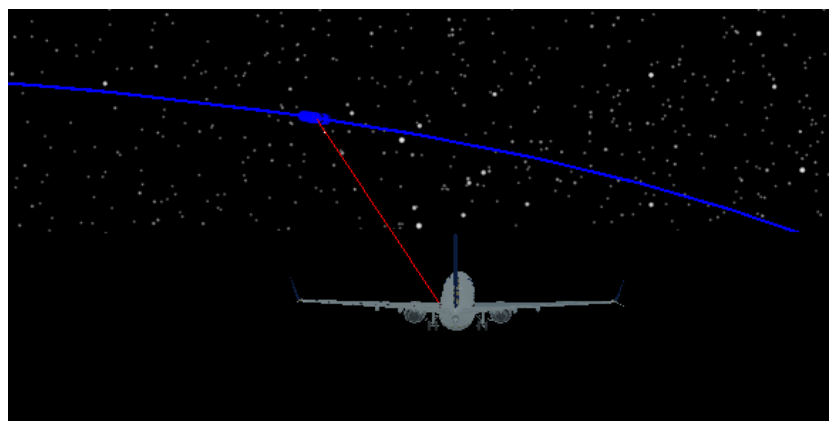

**Figure 11.** Screenshot from STK showing the Starlink satellite train (blue satellite icons) propagated to Aug 10th, 2022, 11:39:08UTC, the time of the first photograph from the perspective of the AC536 aircraft (depicted using a CAD model (Goo 2022)). The path of the satellite train is also blue.

Finally, as the ADS-B barometric altitude values could be off by as much as 2500 feet at altitude (S. Narayanan 2022), we provided Figures 9 to show no significant difference in the simulated satellite train's apparent location with an additional 2500 feet added to the MSL altitude (+0.8 km). Further, looking at the processed values from the Python script output files one notices that the potential for an altitude error leads to negligible relative position differences to the satellites based on how far they are from the aircraft. The screenshots at the time of photo 2 show that the train is almost on top of $\alpha$-Gem (Castor). Additional screenshots showing the locations when an additional altitude error is added to the aircraft also demonstrate negligible change in the relative location of the satellite train for both photos.

**Blender Visualization Results**

Blender 4.0.1 was used in an attempt to create a realistic cockpit view of the Starlink train using ray tracing (Blender 2023). To place the Starlink train in the correct location as viewed from the cockpit, we translated our look angle coordinates into cartesian coordinates centered on the CAD model of a Boeing aircraft. Size estimates of a Starlink satellite were obtained from two different references (reddit users 2023) and (Forrest 2020), neither of which are official SpaceX/Starlink publications. This is an issue but is the best current information that we must work from.[22]

Figures 12 shows a zoomed in of Starlink satellites with the solar arrays in the stowed and deployed (open-book) configuration, respectively. These images also document the relative locations, rotation angles (X-axis is -170 degrees, Y-axis is 31.92 degrees, and Z-axis is -10 degrees), and sizes for the model used in this paper.[23]

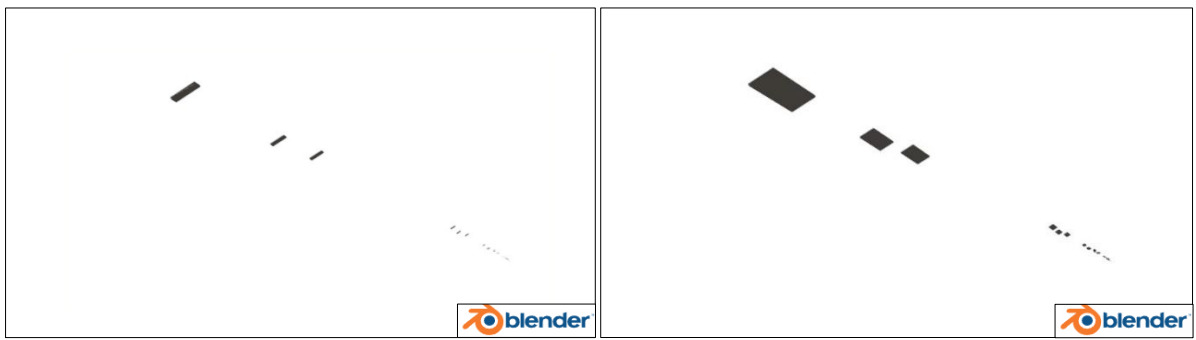

**Figures 12a and b. Blender screengrabs of our simple Blender model of Starlinks using Blender's "Cube" for simplicity. In Fig 12a, represents the satellites with their solar arrays in the stowed for launch configuration, the Cubes are dimensioned to 2.8 meters in width by 1.4 meters in length and 0.2 m in height (or thickness) representing the satellites in the stowed solar array configuration. In Fig 12b, the open-book configuration, the Cube representing each satellite, has had the length dimension modified by adding an additional 10 meters to change the length to 11.4 meters to represent their solar arrays being deployed (the "open-book" configuration) while they are in their orbit boost phase. In both images, Blender camera's viewpoint has been moved from within the aircraft's CAD model to a mere 161.2 meters from the nearest satellite (Starlink-4479) to show the change in observed surface area in this orientation. The orientation was selected to have the solar array extend into the general direction of travel of the satellite train, and the Y-rotation angle is the same angle as the average of the sun's grazing angle from Photo 1's UTC time using all of the satellites as determined by our Python script. This orientation was selected in an attempt to properly mimic**

---

[22] A request sent to SpaceX for Starlink geometry information by one of us went unanswered.
[23] Note that the website contains several additional blender models and their rendering results representing our attempts at experimenting with the various parameters.

**a slightly off "knife-edge" orientation for what we felt is a likely orientation for the satellite based on available information. Each satellite has been placed at the correct relative distance from the AC536 aircraft as were calculated by our Python script from the first photo's UTC time. The subsequent satellites in view are Starlink-#s 4304, 4484, 4476, 4480, 4545, and so forth. From this closer viewpoint, the clustering of satellites as the cause for the observed gaps is easily observed.**

Figure 13 displays the relative location of the satellite train from within the cockpit, not considering local roll-pitch-yaw aircraft dynamics as there was no data to suggest significant perturbations to the ADS-B data for these axes.

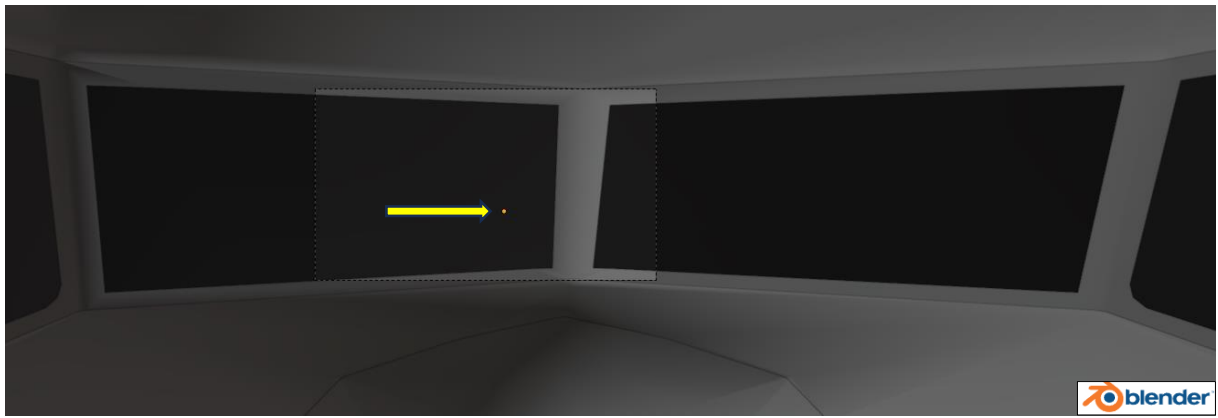

**Figure 13. Blender screengrab showing the location of the Starlink train (the yellow dot just to the right of the yellow arrow) as viewed from the cockpit at the time of photo 1 as determined from ADS-B and Group 4-26 TLE data and transformed into cockpit view coordinates. The grey box surrounding this area of the screen is the location of Blender's camera view, and represents the area that is rendered, as shown below in Figure 14 and Figure 15.**

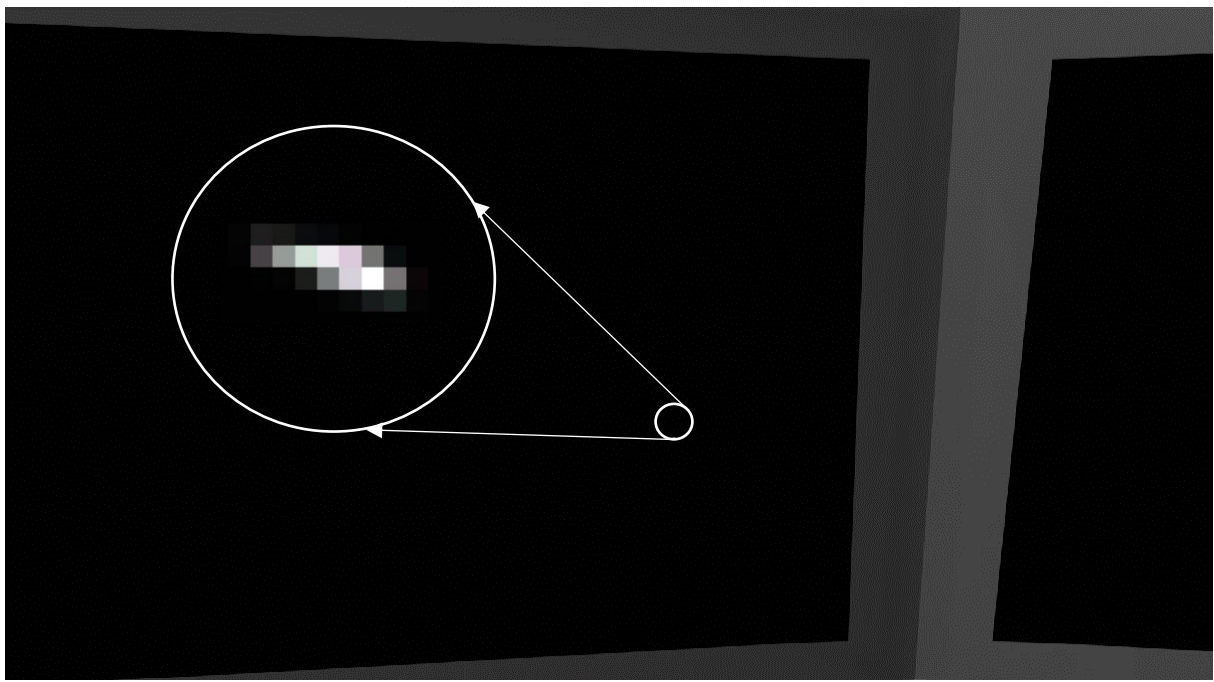

**Figure 14. Blender rendering of the Starlink train in the open-book solar array configuration from one of our rendering experiments. The insert shows a zoomed in view of the rendering with these settings. The rendered result appears to only show two satellites with the remainder**

**hidden behind these two.[24] It took 9 minutes and 50.48 seconds to render this image on the HP Omen laptop that was used for these experiments.[25]**

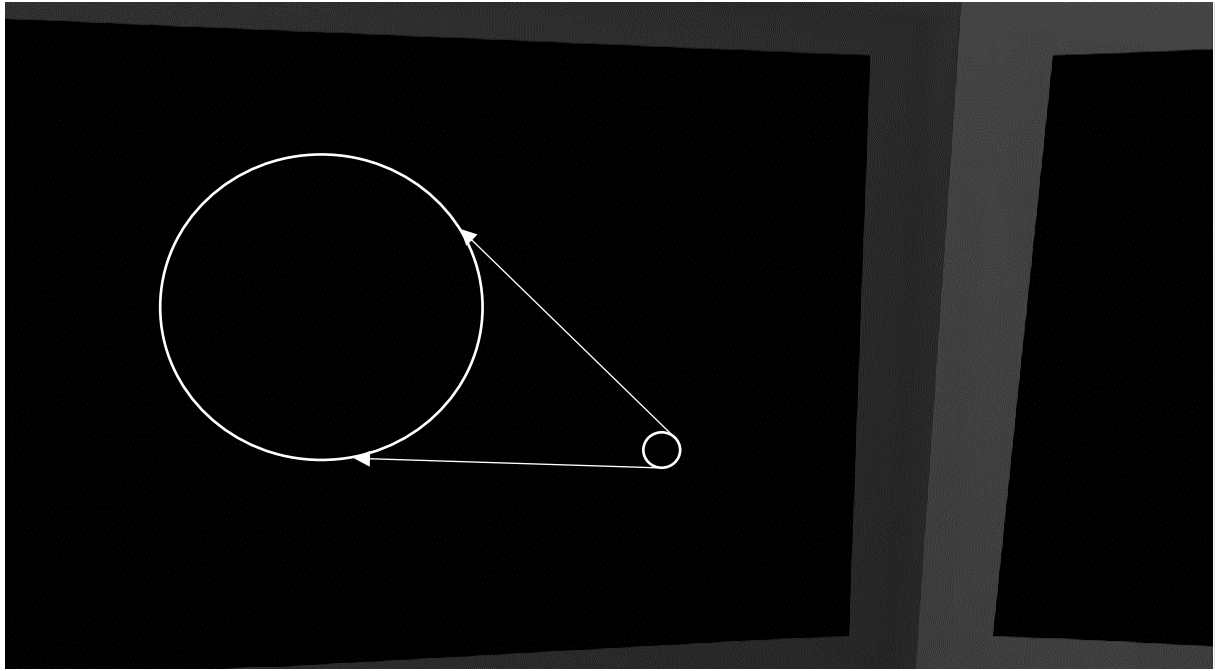

**Figure 15. Blender rendering of the Starlink train in the stowed, undeployed solar array configuration from one of our rendering experiments. The insert shows a zoomed in view of the rendering with the same rendering settings as used in the open-book configuration. Here we see for this experiment that nothing is observed.[26] There are, however, experiments where a very faint object is visible. Due to the time to render each of these images on our hardware, and after rendering several different experiments with variations in Starlink orientation, emission intensity, noise levels, number of samples and others, we felt that we had in principle demonstrated the type of analysis that should be performed. We further concluded that this work should be revisited in the future using specialized space-based rendering software such as that from the OpenSpace Project (OpenSpace Team 2023).**

Figure 14 and Figure 15 are ray-traced Blender renderings of the open-book and stowed (undeployed) Starlink train at the distances observed at the time of photo 1. To do the renderings, Starlinks are configured as emission surfaces, and in both cases had the strength set to a million with 8192 samples.

These experiments, while not definitive, suggest that the solar arrays had to have been deployed for the satellites to have been observed by the pilots in both aircraft. Additional work to better test this result and the effect of other perturbations such as the cockpit's window, and JPEG compression is discussed in the Future Work section of this paper.

---

[24] Details for the rendering settings for this experiment can be found on our GitHub site in the file STLNew_ph1-rotated-open.blend. This effect may in fact be due to the resolution limitations of Blender.
[25] We used a 3.2 GHz AMD Ryzen 7 5800H with Radeon Graphics, and an NVIDIA GeForce RTX 3060 GPU at driver version 546.26 with 7.34 GB of Video RAM.
[26] Details for the rendering settings for this experiment can be found on our GitHub site in the file STLNew_ph1-rotated-closed.blend

**Geometry of the Satellite Train's Observation**

Extracting Earth Centered Inertial (ECI) based state vectors for the aircraft, the satellites, and the sun at the time of the photographs from SOAP is straightforward.[27] We want to do this to compare our geometry to the case study. Figure 16 shows the vector geometry of a Starlink satellite train relative to an aircraft.

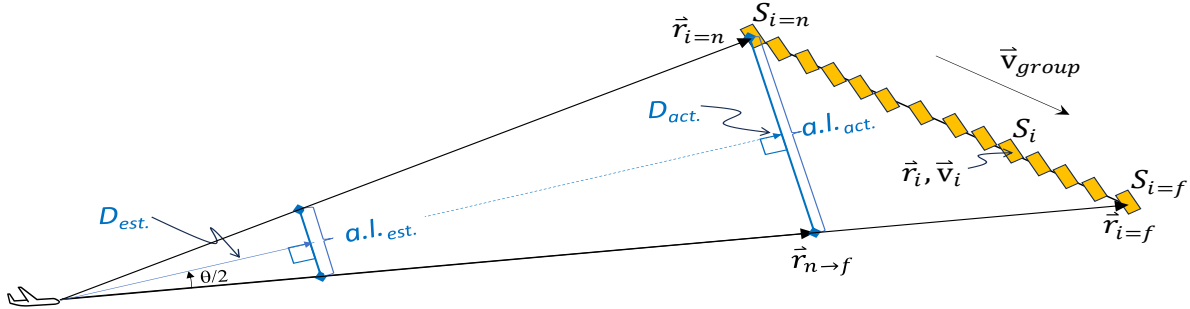

**Figure 16.** This is a vector diagram representation of the location of the $i^{th}$ Starlink satellite in the train with $i = n$ for the nearest visible satellite in the train, and $i = f$ for the furthest visible satellite in the train, as viewed from the cockpit. The diagram is drawn in relation to the common method for calculating the size of a rigid object at an estimated distance $D_{est.}$ and is represented by the solid blue arrow ending at the estimated apparent length $a.l._{est.}$ which is also in blue font. $D_{est.}$ is perpendicular to $a.l._{est.}$. The apparent actual distance, $D_{act.}$ is represented by the continued dashed blue arrow that starts at the aircraft and ends at the actual apparent length $a.l._{act.}$ which is also in blue font. $D_{act.}$ is perpendicular to $a.l._{act.}$. Hence, the actual apparent length starts at the distance from the aircraft to the nearest satellite with state vector $\vec{r}_{i=n}, \vec{v}_{i=n}$ relative to the aircraft and is represented as $S_{i=n}$ in the diagram. The actual apparent length terminates along the line of sight (LOS) to the last visible satellite in the group with the position vector is $\vec{r}_{i=f}$, so that the actual apparent length is perpendicular to the actual apparent distance and is the projection of the nearest visible satellite's position vector onto the furthest visible satellite's position vector $\vec{r}_{n \to f}$. The apparent velocity of the group ($\vec{v}_{group}$) would be moving from the upper left to the lower right in the field of view. The $i^{th}$ Starlink satellite is $S_i$ with state vector $\vec{r}_i, \vec{v}_i$. The nearest satellite is represented as $S_{i=n}$, and $S_{i=f}$ depicts the furthest satellite. Likewise, the nearest Starlink's position is $\vec{r}_{i=n}$, and the furthest satellite's position vector is $\vec{r}_{i=f}$. The angle $\theta$ between these LOS position vectors is the actual apparent length in degrees.

The object's estimated apparent length in degrees (the angle $\theta$) was estimated from the photo to be about 1.5 degrees, with a pilot-estimated distance $D_{est.}$ of ~20 to 30 miles. The apparent length $a.l._{est.}$ at these distances was estimated to be about a mile at 30 miles in distance (1.61 km) in the case study's photometry section. The actual apparent length is identified in the diagram as $a.l._{act.}$, where the value for the actual apparent length in degrees is calculated from,

$$\cos\theta = \frac{\vec{r}_n \cdot \vec{r}_f}{|r_n||r_f|} \Rightarrow$$

$$\theta = \cos^{-1}\left(\frac{\vec{r}_n \cdot \vec{r}_f}{|r_n||r_f|}\right)\frac{180^o}{\pi}$$

Note that for the aircraft, we have a 'Smooth Route' option selected, where SOAP employs an algorithm that "banks the aircraft" between route points, rather than performing abrupt changes in direction.[28] Aircraft relative state vectors for each satellite are then computed as,

$$\vec{r}_{i\ rel\ to\ ac.} = \vec{r}_{i\ in\ ECEF} - \vec{r}_{ac.in\ ECEF}$$
$$\vec{v}_{i\ rel\ to\ ac.} = \vec{v}_{i\ in\ ECEF} - \vec{v}_{ac.in\ ECEF}$$

---

[27] This is a "copy to clipboard" option available within SOAP when displaying each platform's (sun, aircraft, or satellite) data view, which were each saved to a different ascii text file.

Recalling that the magnitude of a vector is given by,

$$|r| = \sqrt{r_x^2 + r_y^2 + r_z^2}$$

From these extracted state vectors, in addition to calculating the apparent lengths (*a.l.*$_{act.}$ and *a.l.*$_{est.}$) using Earth-Centered-Earth-Fixed (ECEF) coordinates, we wished to convert the coordinates to be relative to the aircraft. We first tried various coordinate systems for the aircraft commonly used for satellites, such as NTW (Nadir-Track-Wing) coordinates and RSW (R is radial direction pointing away from the Earth, S is along track, and W is the cross-track direction), settling on defining an ENU (East-North-Up) based cockpit relative representation, as described below. NTW was however still however appropriately used for calculating the sun grazing angle where the N-axis lies in the orbital plane pointed at the Earth, T is tangential to the orbit along the velocity vector, and W is normal to these axes.[29]

Hence, based on the geometry in Figure 16 above, and Figure 17 below, we derive the sun's grazing angle and the apparent length of the satellite train. To do this we transform ECEF coordinates for the aircraft and the sun to be relative to each of the Starlink satellites in NTW. We then calculate the angle between each of these vectors, using the same vector dot product equation for ($\theta$) above. The grazing angle ($\varphi$) in degrees is then given by,

$$\varphi = (180 - \theta)/2.$$

Figure 17 uses the same diagram format found in (Fankhauser, Tyson and Askari 2023) to show the grazing angle with the angles between the sun and the aircraft relative to the satellite.

---

[28] It is possible this option can introduce variations in the relative position of the satellites as viewed from the cockpit with ADS-B data that is not updated say on the order of every minute. Hence, we did experiments with this option disabled as well. It does add measurable changes to the state vector of the aircraft.

[29] Note that we tried calculating the grazing angle using both NTW and RSW coordinates where RSW is like the roll-pitch-yaw coordinate system (RPY), where R→yaw, S→roll, and W→pitch. As there was no difference in the grazing angle results, we elected to keep just the NTW calculations in our script. For a description of the difference between these satellite coordinate systems, we refer the reader to (Vallado 2013) pgs. 155-158 for more information.

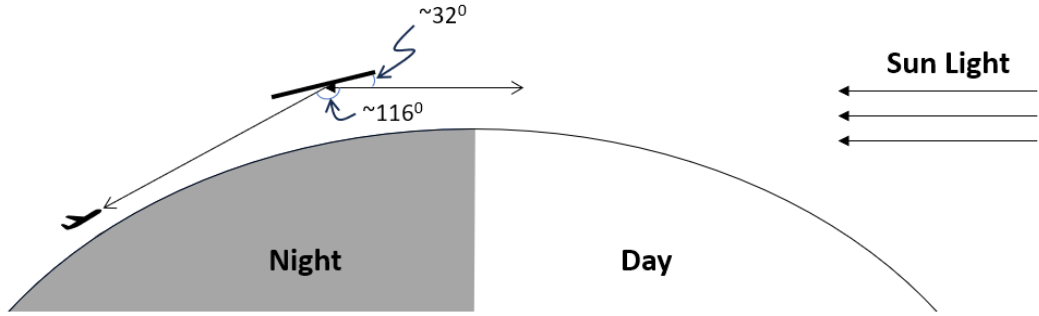

**Figure 17. Geometry showing the geometry of the specular reflection observed by AC536 Photo's 1 and 2. This is a modified version of Figure 1 from (Fankhauser, Tyson and Askari 2023) using the average rounded up sun grazing angle for both photos in the image (~32 degrees). The original Figure 1 also includes the Earth shine component that should be included, however, at this time we feel requires more advanced rendering tools. It is also important to note that our geometry and the brightness of the Starlink satellites strongly suggests that specular reflection is the predominant reflection component for our case study geometry. To properly confirm this hypothesis, we would need to know the actual attitude (orientation) of each satellite and be able to incorporate the reflective properties of the satellite's surface materials into a CAD model. We will discuss this further in later sections of the paper.**

To simulate the location of the Starlink train in the cockpit, we also needed to transform ECEF from SOAP into aircraft heading relative ENU coordinates. We define the "look angle" as positive from the heading clockwise and negative counterclockwise from the aircraft's heading. The elevation angle is defined to be positive in the "up" direction and negative in the "down" direction. Absent information regarding the real-time roll-pitch-yaw status of the aircraft we assume these perturbations are non-existent in the following derivation of the transformations from ECEF to ENU with a heading angle rotation correction.

To transform ECEF into local ENU coordinate system at the aircraft's location we use the aircraft's latitude ($\phi$), and longitude ($\lambda$). The altitude of the aircraft is embedded in the ECEF coordinates extracted from SOAP. The transformation from ECEF to ENU coordinates involves rotating first about the Z-axis by ($-\lambda$) to align the prime meridian with the local meridian, then about the transverse axis by ($\frac{\pi}{2} - \phi$) to align the up direction with local vertical. The resulting rotation matrix to ENU is,

$$R_{ECEF \rightarrow ENU} = R_\phi R_\lambda = \begin{bmatrix} -\sin\lambda & -\sin\phi\cos\lambda & \cos\phi\cos\lambda \\ \cos\lambda & -\sin\phi\sin\lambda & \cos\phi\sin\lambda \\ 0 & \cos\phi & \sin\phi \end{bmatrix}.$$

Each of the columns of this rotation matrix define the three-unit vectors $\hat{e}$, $\hat{n}$, and $\hat{u}$ respectively. Following the ECEF to ENU conversion we then need to account for the aircraft's heading to provide a cockpit view. We rotate the East and North axes by the *negative* heading angle to account for our desire to have the look angle be positive in the clockwise direction and negative in the counterclockwise direction.

$$R_{heading} = \begin{bmatrix} \cos(-\theta_{heading}) & \sin(-\theta_{heading}) & 0 \\ -\sin(-\theta_{heading}) & \cos(-\theta_{heading}) & 0 \\ 0 & 0 & 1 \end{bmatrix} \Rightarrow$$

$$\begin{bmatrix} \cos(\theta_{heading}) & -\sin(\theta_{heading}) & 0 \\ \sin(\theta_{heading}) & \cos(\theta_{heading}) & 0 \\ 0 & 0 & 1 \end{bmatrix}$$

Finally, the look angle ($\alpha_{look}$) and the elevation angles ($el$) in degrees are calculated from the following equations,

$$\alpha_{look} = ArcTan2(E_{ECEF \to ENU \to heading}, N_{ECEF \to ENU \to heading}) \frac{180}{\pi}$$

$$el = ArcTan2(U_{ECEF \to ENU \to heading}, D_{horizontal}) \frac{180}{\pi}$$

with $D_{horizontal}$ given by,

$$D_{horizontal} = \sqrt{E_{ECEF \to ENU \to heading}^2 + N_{ECEF \to ENU \to heading}^2}$$

To obtain these results we wrote three different Python scripts (with the support from OpenAI's ChatGPT), the first of which translates the SOAP Starlink ephemerides that were copied into ASCII files, into comma-delimited (csv) files. The second script then processes each of the csv files to methodically calculate for each satellite and aircraft file; the sun grazing angle, the apparent length, and the heading corrected locations. A third python script contains the rotations from (Vallado 2013). Table 1 provides a summary of our results from processing the SOAP extracted data from both photos.

**Table 1. Summary of results from processing SOAP simulation extracted data with the route smoothing algorithm off. The *output-analysis.xlsx* excel spreadsheet provided on our GitHub site (see section 9 later in this paper) also includes data from when this SOAP algorithm was enabled.**

| SOAP Simulation Option: Name (units) | Photo 1 at 11:39:08UTC | Photo 2 at 11:39:24UTC |
|---|---|---|
| **SOAP route smoothing off** | | |
| Apparent Length (deg) | 1.12 | 1.05 |
| Actual Apparent Length (km) | 30.78 | 29.8 |
| Actual Apparent Length (mi) | 19.13 | 18.52 |
| *Estimated Apparent Length @ 30 miles (mi)** | *0.59* | *0.55* |
| Average Grazing Angle (deg) | 31.92 | 31.95 |
| Nearest Satellite Distance (km) | 1574.4 | 1626.2 |
| Nearest Satellite Distance (mi) | 978.5 | 1010.69 |
| Furthest Satellite Distance (km) | 1590.5 | 1644.1 |
| Furthest Satellite Distance (mi) | 988.5 | 1021.81 |
| Max Look Angle (deg) | -7.42 | -3.75 |
| Min Look Angle (deg) | -8.53 | -4.79 |
| Average Look Angle (deg) | -7.93 | -4.23 |
| Max Elevation Angle (deg) | 2.41 | 1.9 |
| Min Elevation Angle (deg) | 2.24 | 1.72 |
| Average Elevation Angle (deg) | 2.32 | 1.8 |
| Distance Traveled from Photo 1 to 2 (deg) | 3.74 | |
| Tangential Speed Photo 1 to 2 (deg/s) | 0.23 | |
| * The values in this row are estimated lengths based on the pilot's perceived distances. | | |

**Table 2. Summary of results from processing SOAP simulation extracted data with an additional 2500 feet (0.8 km) added to the altitude to simulate an error in the aircraft's altitude. Differences from the addition of the altitude error are shown below for each photo and simulation parameter.**

| SOAP Simulation Option: Name (units) | Photo 1 at 11:39:08UTC | % Diff. | Photo 2 at 11:39:24UTC | % Diff. |
|---|---|---|---|---|
| Apparent Length (deg) | 1.12 | 0.00 | 1.05 | 0.00 |
| Actual Apparent Length (km) | 30.7 | 0.26 | 29.74 | 0.20 |
| Actual Apparent Length (mi) | 19.08 | 0.26 | 18.48 | 0.22 |
| *Estimated Apparent Length @ 30 miles (mi)* | *0.59* | *NA* | *0.55* | *NA* |
| Average Grazing Angle (deg) | 31.93 | -0.03 | 31.97 | -0.06 |
| Nearest Satellite Distance (km) | 1570.6 | 0.24 | 1622.6 | 0.22 |
| Nearest Satellite Distance (mi) | 976.13 | 0.24 | 1008.45 | 0.22 |
| Furthest Satellite Distance (km) | 1586.7 | 0.24 | 1640.6 | 0.21 |
| Furthest Satellite Distance (mi) | 986.14 | 0.24 | 1019.64 | 0.21 |
| Max Look Angle (deg) | -6.66 | 10.24 | -2.94 | 21.60 |
| Min Look Angle (deg) | -7.77 | 8.91 | -3.98 | 16.91 |
| Average Look Angle (deg) | -7.17 | 9.58 | -3.41 | 19.39 |
| Max Elevation Angle (deg) | 2.42 | -0.41 | 1.91 | -0.53 |
| Min Elevation Angle (deg) | 2.25 | -0.45 | 1.73 | -0.58 |
| Average Elevation Angle (deg) | 2.33 | -0.43 | 1.81 | -0.56 |
| Distance Traveled from Photo 1 to 2 (deg) | | | 3.8 | -1.60 |
| Tangential Speed Photo 1 to 2 (deg/s) | | | 0.24 | -4.35 |

Using the location of the Starlink train we also plotted in Microsoft Excel® each satellite's look angle and elevation angle against the Gemini star background at the UTC times of both photographs (Figure 18 and Figure 19) as an additional visual check on the location of the Starlink satellite train. These data were also used to calculate the angular distance traveled between the photographs. The result was 3.7 deg in 16 seconds.

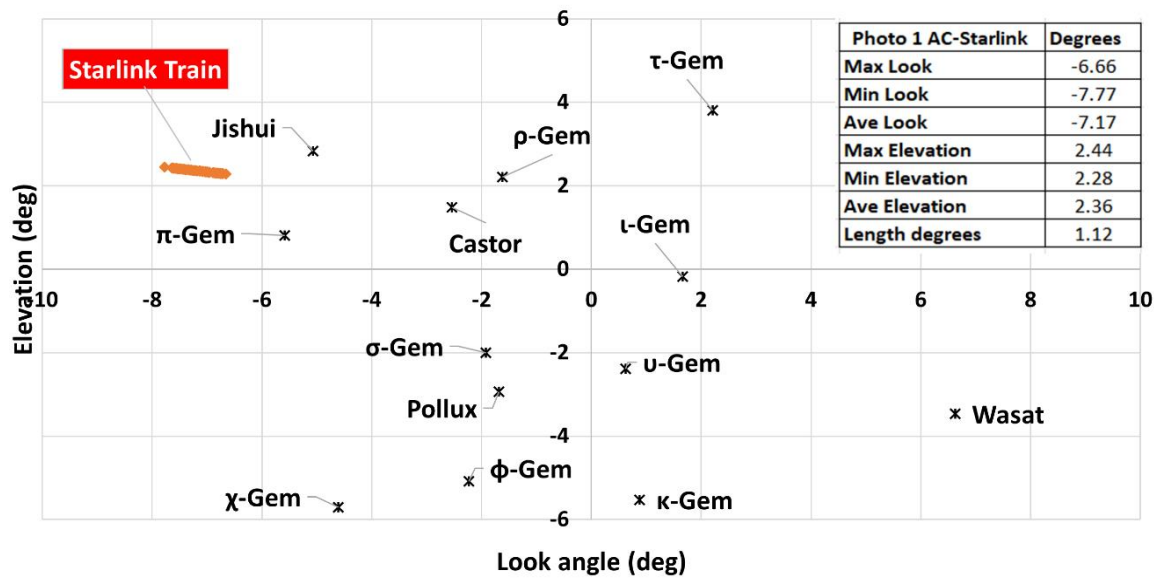

**Figure 18. Microsoft® Excel® plot of the derived look and elevation angles for the satellite train at the time of Photo 1 with prominent Gemini stars included. [30] Agrees with location in photograph.**

---

[30] We used Mathematica® version 13.2 "ConstellationData" function to identify the Right Ascension and Declinations for bright stars in Gemini. These were converted into decimal degrees using Mathematica's "FromDMS" function and rotated into aircraft look and elevation angle coordinates. The Mathematica

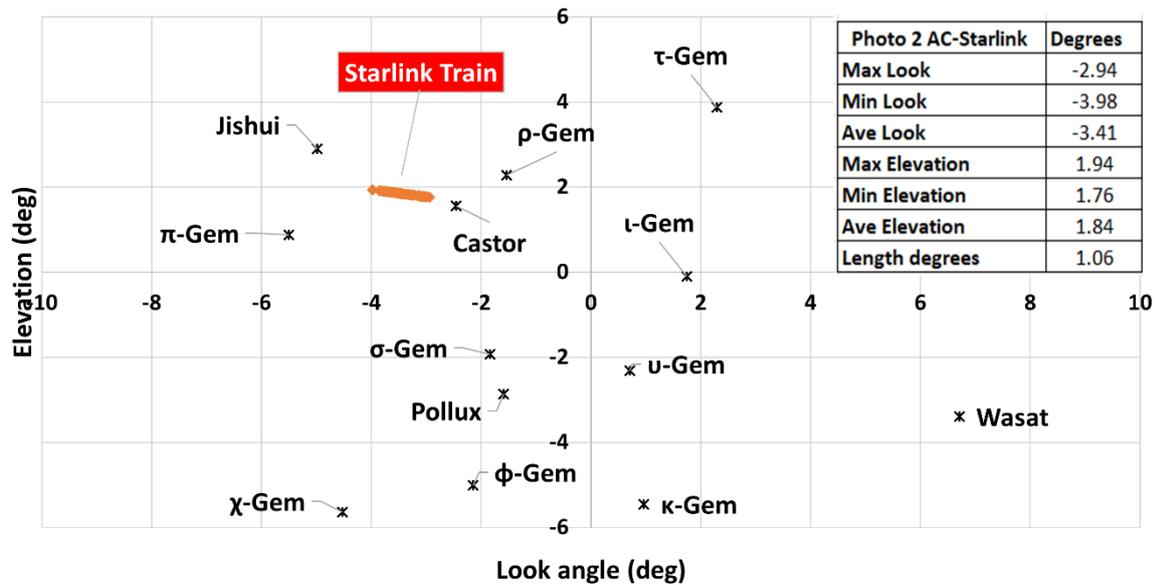

**Figure 19.** Microsoft® Excel® plot of the derived look and elevation angles for the satellite train at the time of Photo 2 with prominent Gemini stars included. Agrees with photograph.

The model predicts that the object would have a tangential angular speed of about 0.23 deg/s as viewed from the airplane, in close agreement with the case study results of 0.26 deg/s. We describe the potential causes for this discrepancy in the next section.

## 4. Discussion

Our analysis strongly suggests that this unidentified object is likely from an earlier SpaceX launch on August 10th, 2022, of version 1.5 Starlink satellites. The satellite train was crossing the day/night terminator at about 1620 km in distance from Air Canada flight (AC536) from which an anonymous pilot took photographs. Table 3 provides a comparison of the case study derived observables to the equivalent geometry derived values.

**Table 3. Comparison of Case Study Derived Observables to Similar Geometry-based Values**

| Name (units) | Case Study | Geometry-based |
|---|---|---|
| Apparent Length (deg) | ~1.5° | ~1.1° |
| Apparent Speed (deg/s) | 0.26 | 0.23 |
| Apparent magnitude | -4 | Incomplete[31] |

When closely looking at Figure 3 we see JPEG compression artifacts (Wikipedia 2024b) in the iPhone 12 photograph. Understandably the default image collection

---

Notebook and the Python function stars.py which converts far more stars into aircraft coordinates than are plotted are both available on our GitHub site.

[31] Deciding that Blender is not the correct tool to properly estimate this quantity, we've decided to defer an attempt to render the apparent magnitude for future work using OpenSpace to include atmospheric scattering and other image perturbing artifacts such as cockpit window and camera optics and compression effects. In addition, while we prefer having the actual attitude (orientation) for each satellite for further analysis, one could in principle with sufficient computing power create a detailed test matrix of orientations or a Monte Carlo analysis to help identify the best match to what was observed by the pilots. Subsequent discussions, (Mallama 2024), indicates our apparent magnitude is consistent with their estimation for a single spacecraft.

mode for most cell phone cameras, which regrettably removes fine details from an image that may have supported a more rapid identification of the source of the sighting. Further, the images in Figure 1 and Figure 2 clearly show atmospheric scattering from the sun, which is observed as the change in coloration of the background above and below the word "Castor". It is entirely possible that the images of the Starlink train may have also been distorted by this phenomena. Hence, we suggest that both perturbing factors could account for the differences between our image based apparent length and the geometry based apparent length that simply modeled these satellites as a point at the SGP4 propagated location.

As discussed earlier, the impact of Starlink satellites on astronomy has been known for some time (McDowell 2020, Girgis 2019, Cui and Xu 2022) as has their misidentification as UAP/UFOs (CBS Pittsburgh 2021, WCVB Channel 5 Boston 2023, Reyes 2023, Tangermann 2019, Grassi 2023, Mandelbaum 2019). After the impact on observational astronomy became known, SpaceX/Starlink and the astronomical community have been working to reduce the optical signatures of constellation satellites, with improvements to their next generation of Starlinks (v2.0 minis and v2.0) (Tangermann 2019), (Loeffler 2023), (SpaceX 2022), (S. S. Committee 2020), (S. S. Committee 2021), (Crider 2022). Froust at SPACENEWS has documented the ongoing history of this issue in these articles (Foust 2020a, Foust 2023). In addition to Fankhauser, Tyson and Askari 2023, which characterizes how these airlines in the pacific were able to so clearly see these Starlinks, there are additional papers regarding the optical characteristics of Starlinks (Mallama, Hornig, et al. 2023, Mallama, Cole, et al. 2023). Private communications with researchers suggest that they are continuing to actively investigate SpaceX's attempts at mitigating the visibility of these satellites with more research results forthcoming (Mallama 2024, Cole 2024).

Importantly, a key paper published on Starlink's API website (SpaceX 2022) includes a log-linear plot of the 'bidirectional reflectance distribution function' (BRDF) for the dielectric mirrors for the various generations of satellites, Figure 20. However, we found attempting to incorporate this data directly into Blender to be problematic and CAD models were only available from hobbyists and not well suited for Blender's rendering environment.

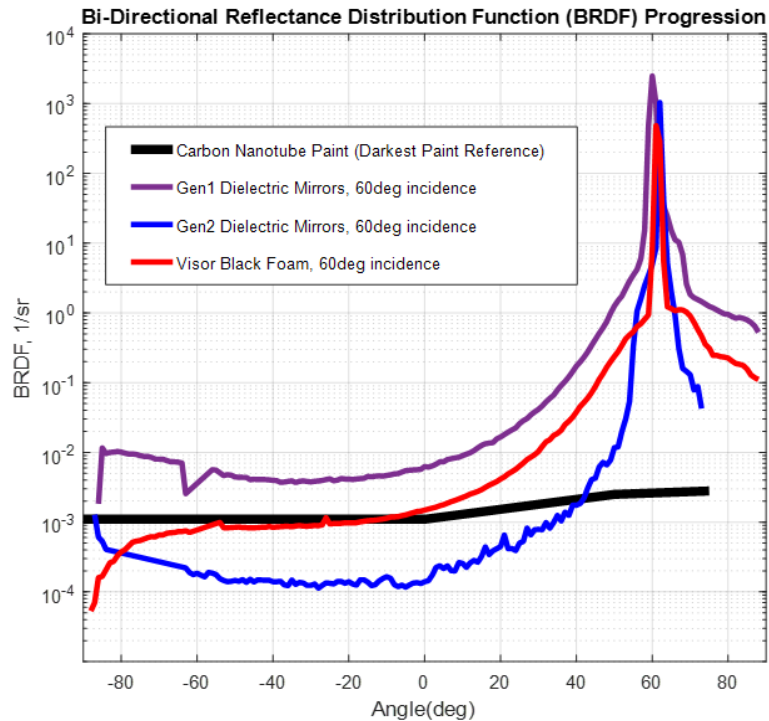

**Figure 20. The Progression of the 'bidirectional reflectance distribution function' (BRDF) from two generations of Starlink satellite dielectric mirrors. Note the significant specular reflection spike at 60 degrees with an incident angle from the light source of 60 degrees.**

**Source: SpaceX (SpaceX 2022)**

The incident angle for the light source is at 60° which is equal to the reflection peak's angle. Therefore, the 2nd generation of Starlink satellites continue to have significant specular reflections (Wikipedia 2023b) from the same angle of incidence. In addition, this geometry is similar to the BRDF angle that we calculated from the sun's grazing angle (see Table 1), i.e., the Sun's grazing angle and the line-of-sight (LOS) angle from the satellite to the aircraft are nearly equivalent. Hence, we do not expect the new satellites to appreciably improve this issue while they are in the orbit boost phase with their solar arrays opened. SpaceX/Starlink appears to be fully aware of the specular reflection issue during their boost phase (Musk 2020). As Musk indicated, one option for dealing with the specular reflection peak is to re-orient the satellites during the orbit boost phase as they are crossing the day/night terminator (see for example pg. 6 in (Musk 2020) and (SpaceX 2020)). This reorientation of their "open-book" configuration would not have the solar arrays and satellite "flat to the earth" and would instead redirect the sun's light back out into space. In (SpaceX 2020), this reorientation is described as, "rolling the satellite so the vector of the Sun is in-plane with the satellite body, i.e. so the satellite is knife-edge to the Sun."

Another would be to include a sunshade that is deployed to reflect the sun back into space when in this orientation. However, knowing that SpaceX had experimented with a "sun visor" option (Foust 2020b), all we really know is that this option was abandoned because it would not solve astronomy's issue and introduced other operational problems (McFall-Johnsen 2020).

As a very recent Starlink train of 22 v2.0 minis (Schrader 2023) was uploaded to YouTube (LIVE 2023) clearly showing the train of Starlinks as they transited across the day/night terminator. This suggests that a re-orientation knife-edge attitude maneuver has yet to be implemented. In addition, researchers who measure the brightness of Starlinks suggest that they do not use the "knife-edge" re-orientation until they are above an altitude of about 360 km (Mallama 2024).

Further, upon closely reading (SpaceX 2020) we find the following additional information about SpaceX's use of the knife-edge to the Sun orientation,

"This would reduce the light reflected onto Earth by reducing the surface area that receives light. This is possible when orbit raising and parking in the precession orbit because we don't have to constrain the antennas to be nadir facing to provide coverage to internet users. However, there are a couple of nuanced reasons why this is tricky to implement. First, rolling the solar array away from the Sun reduces the amount of power available to the satellite. Second, because the antennas will sometimes be rolled away from the ground, contact time with the satellites will be reduced. Third, the star tracker cameras are located on the sides of the chassis (the only place they can go and have adequate field of view). Rolling knife edge to the Sun can point one star tracker directly at the Earth and the other one directly at the Sun, which would cause the satellite to have degraded attitude knowledge.

There will be a small percentage of instances when the satellites cannot roll all the way to true knife edge to the Sun due to one of the aforementioned constraints. This could result in the occasional set of Starlink satellites in the orbit raise of flight that are temporarily visible for one part of an orbit."

## 5. Conclusions and Recommendations

We described a methodology that uses orbital simulation software with aircraft ADS-B location data and a satellite's ephemeris data from supplemental versions of NORAD's Two-Line Elements (TLEs). We initially used The Aerospace Corporation's Satellite Orbital Analysis Program (SOAP) to verify the locations of STARLINK Group 4-26 satellites, but as this software is only available to employees of Aerospace and their U.S. government customers, we also enlisted the support of students enrolled in the University of Utah's Department of Mechanical Engineering's Space Mission Engineering course to model the Starlink train's location in orbit using the commercial orbital analysis package, System Tool Kit (STK) from Ansys.

Images extracted from both simulations clearly show that deploying satellites in Starlink Group 4-26 were in the same location of the sky at the time the photographs were taken. Our analysis also suggests that there are significant information gaps, image compression, and software visualization issues that kept us from providing identical renderings of the satellites as seen by the images in the photographs. Where in principle, physics-based rendering should provide photo-realistic visualization of the satellites during orbit boost if models for all perturbing factors can be developed and compared to the original JPEG compressed iPhone images.

We should also point out a study that found the accuracy of TLEs for the purpose of modeling Starlinks using NORAD provided TLEs provided by Space-Track.org predominantly had an along-track error and were accurate to within 0.3±0.28 seconds with a high value of 1.2 seconds (Halferty, et al. 2022). Thus, attempting to use supplemental SGP4 propagated TLEs very early in the launch sequence prior to obtaining NORAD's RADAR appears viable. These TLEs are from Celestrak's working with SpaceX, as we've described, and is based on our conversations with Kelso (Kelso, private communication 2023).

Reviewing our modeling difficulties, we had to make assumptions regarding the deployment status and orientation of the satellites at the time of the photographs. In addition, the case study suggested the star-like objects merged into a single larger craft-like object, but no photographic evidence was provided. This situation could be explained by changes in the viewing geometry of the satellite's attitude. The visibility of the satellites, as our modeling suggests is dependent on the deployment of the solar arrays, hence, we consider knowing the status of each of the satellite's solar arrays with respect to the satellite's attitude and orbit location over time as critical information. Furthermore, allowing pilots to routinely provide photographic evidence from either an automated system when they are busy with safety critical aircraft operations, or using cell phones when they are not allowed, would significantly improve analysis of UAPs.

In addition, our approach attempted to couple orbital modeling with rendering using Point of View (POV) ray tracing methods to simulate the appearance of a satellite train for a viewer in the cockpit of the aircraft that had photographs. However, including our inability to directly incorporate the reflective material properties on a verified CAD model of these satellites, we also found it difficult using the selected software to provide dependable rendering results using the sun as the illumination source. This led us to simply render the satellites as rectangular emission sources at the correct distances of each satellite to the aircraft. A potential remedy for this difficulty is discussed further in the Future Work section of this paper. However, the results suggest that the satellite train likely had their solar arrays deployed making them visible to the pilots due to the additional area with specular reflection of the sun directed at these aircraft as our hypothesized phenomena.[32] Further, reading carefully (SpaceX 2020), we find that they conclude this observation problem for observers on the Earth will persist.

Therefore, we recommend that satellite operators either unilaterally provide planned spacecraft/satellite attitude orientation, orbit location and deployment timelines, or the government require that they provide this critical information. This information should allow us to provide model-based visualization results for nominal launch timelines for what the spacecraft would look like to observers on the ground or in aircraft with sufficient computing power. Especially for ground-based observers, weather/cloud coverage predictions would be necessary for observability predictions. These nominal models could be updated should unplanned launch delays occur within acceptable tolerances to viewing conditions, potentially updated with sufficient computing power, or notices to aviators and the public could be updated or cancelled.

Finally, the Federal Aviation Administration (FAA) could reduce the risk to the flying public from pilots engaging in airwave chatter (Hansen 2024) trying to figure out what they are seeing by placing this modeling capability in the hands of air traffic controllers, after carefully thinking through a system and communication architecture that provides this additional information for them to inform the pilots. We envision this as a collaborative effort between the satellite's owners (in this case SpaceX/Starlink), the FAA, NASA, NOAA, and the U.S. Space Force.

To characterize the unknown, we have to first develop methods to characterize the known.

Those of us in the scientific community must be provided with high-quality reports to develop these methods. Pilots have a unique vantage point aloft that ground observers lack, they are trained in aerial observation, and they have immediate access to important and automated reporting data such as latitude, longitude, altitude, bearing, time, sometimes ground radar, and there are usually two of them. Anything that looks unusual is interesting and useful for improving our overall situational awareness, as we have done here for the UAP report data from the pilots of AC536. We look forward to pending congressional legislation to protect pilots when reporting what they witness in the air (Thomas 2024), preferably allowing them to provide rapid reporting and dissemination of cell phone data. Further, we also look forward to the deployment of multi-wavelength sensors on aircraft that can routinely provide a wide range of additional information about UAPs to further our investigation of the phenomena (Vincent 2023).

---

[32] The primary reflective area of the Starlink satellite would increase by ~8x with solar arrays deployed.

Finally, our preferred list of required information (where a ✓ indicates the data or information exists, and ✘ indicates that either the data or information does not exist, we are not aware of its publicly accessible existence, or it is not usable in its current form) for each launch of satellites would include:[33,34]

- the launch window in Coordinated Universal Time (UTC)
    - ✓ Spaceflight Now (Spaceflight Now 2024)
    - ✓ NASA Space Flight (NASA Space Flight 2024)
    - ✓ RocketLaunch.Live (RocketLaunch.Live 2024) and others
    - ✓ predictive supplemental TLE's from Celestrak and Space-Track.org
- orbit boost timelines for each satellite deployed into a low Earth orbit
    - ✘ predictive TLE's provided to Celestrak and Space-Track.org
- launch debris orbit decay predictions
    - ✓ TLE's from Celestrak and Space-Track.org
- attitude and booster de-orbit ignition dynamics
    - ✘ orientation and ignition plans/models for de-orbit of rockets and their debris (not provided)
    - ✘ orientation plans/models satellites with deployment (solar arrays, antennas) timelines (not provided)
- other items required to properly model these events
    - ✘ dynamic Computer Aided Design (CAD) models of the exterior of the spacecraft with materials and their reflective properties (e.g., BRDF incorporated directly into these models) (not provided directly, however, some approximate hobbyist models without surface properties are available)
    - ✓ weather forecast - in particular cloud coverage at various altitudes to support ground observability models, (NOAA)
    - ✘ software and algorithms to accumulate the above to provide simulations to aviators and ground observers - observability predictions (use STK,

---

[33] Potential national security restrictions could require some launches to remain obscured in secrecy.
[34] We use a ✓ to indicate that a known information source exists, and ✘ to indicate that no known information source exists, or that this is a capability that has not yet been developed.

- ✖ Stellarium[35] or other software, where additional examples are NASA Worldwind (NASA 2023), and Google Earth (Google LLC 2023))
- ✖ software to provide renderings for aircraft and ground observers (exploring OpenSpace is a research goal)
- ✖ sufficient computational resources to provide results in a timely manner (use a cloud-based computing provider)
- ✖ routine sighting reporting/identification software for cell phones and tablets with their allowed use by pilots (some cellphone-based apps exist, such as SIGHTER.io (SIGHTER 2023) and Enigma Labs, LLC (ENIGMA 2023) which in lieu of an aircraft integrated and automated capability could be allowed when not engaged in safety critical operations, but as of the time this paper was released, neither of these services provide predictive capabilities for known aerospace phenomena, and allowing pilots to provide photographic evidence remains problematic)

As stated earlier, this approach should help remove a possibly significant fraction of known space-related events from being mischaracterized and misreported as UAP with estimates ranging from a low of 18% (MUFON) to 25% (NUFORC) of all reports (Cockrell, Murphy and Rodeghier 2023) to perhaps 36% or better using a conservative estimate of the Powell FOIA request (Powell 2024, DrDougB 2024).

## 6. Future Work

Investigating alternative rendering software to include potential image perturbing effects is the most obvious capability we (or others) need to investigate in future work. Specifically, the observability of the deployment of these constellations with respect to a satellite's surface material's reflective properties from variations in solar, earth and lunar illumination angles from satellite surfaces is an important next step. We believe OpenSpace to be the right tool. Appropriate software should provide renderings that can include physics-based atmospheric scattering, cockpit window distortion models including those from random insect-splatter, and cell-phone camera acquisition and storage using more efficient ray tracing algorithms for our modeling situation. Analyzing a full test-matrix for various image perturbations would provide interesting results.

Finally, we anticipate that a properly engineered solution for the FAA would allow ground controllers to provide notices to aviators regarding the visibility of satellites. Eventually, considering the progression of Moore's Law (Wikipedia 2024c), we anticipate the ability to provide tablet or cellphone simulation results to the aviation community and the public would help provide advanced notification for what these space events should look like under specific viewing conditions. However, we feel to do this modeling correctly, and in a timely manner our simulation models would need the information provided in the Conclusions and Recommendations section of this paper.

---

[35] Stellarium is astronomical software that can plot satellites with existing apps that can be downloaded onto both iOS and Android based phones (Stellarium contributors 2023, Stellarium Labs 2024a, Noctua Software Ltd 2024b).

## 7. Acknowledgments

The authors would like to thank the pilots of AC536 for the photographs and their eyewitness accounts of the sighting, as well as the pilots of AC34 for their corroborating eyewitness accounts without which this paper would not have been possible. We are grateful to Mick West of Metabunk.org for recognizing that the object seen by the aircraft pilots was a train of recently-launched Starlink satellites and not a UAP. We would also like to thank Dr. T.S. Kelso for the query to extract all the Starlinks from Group 4-26 supplemental TLEs and Dr. Jonathan McDowell for providing relevant supplemental TLEs and information during the initial research phase of this work. We would like to thank Ben Hansen for his insights into the flight safety issue these reports are causing for the aviation industry, and Dr. T.S. Kelso's support with supplemental TLEs and his review comments. We especially want to thank the Scientific Coalition for UAP Studies science advisor, Dr. Sarah Little, for ADS-B data for the flights and several extremely helpful review comments and suggestions that have been provided during this research. Richard Griffiths acknowledges the support of MUFON for access to the observational sighting report and for discussions relating to the nature of the observed images and video. We thank Anthony Mallama and Richard Cole for insightful discussions regarding their on-going work measuring SpaceX/Starlink reflections. Finally, we also thank an anonymous reviewer for their critical comments resulting in our including several clarifications.

## 8. Conflict of Interest Statement

Two of the authors (Buettner and Griffiths) are consultants for SIGHTER.io, one of the known cell phone apps created to crowd source observations of possible UAP sightings.

## 9. Supplementary Materials

We list here our supplementary materials and methods archived online.

1. A GitHub website with materials used in this research and screenshots of SOAP modeling results is available at (DrDougB 2024):

    **In the "root" directory:**
    i. README.md: standard GitHub description of the archive
    ii. 1st_photo.zip: is a zip archive of the SOAP extracted ephemerides and the csvout processed comma delimited format files for SOAP at the same UTC time as photo 1 was taken. The ouput.csv file is the final processing from the cockpitview.py python script for photo 1.
    iii. 2nd_photo.zip: is a zip archive of the SOAP extracted ephemerides and the csvout processed comma delimited format files for SOAP at the same UTC time as photo 2 was taken. The ouput.csv file is the final processing from the cockpitview.py python script for photo 2.
    iv. AC536.xlsx: AC536 ADS-B data processed to extract location data for SOAP.
    v. AC536_2d008340_alice_rg.csv.kmz: AC536 ADS-B data provided by the Scientific Coalition for UAP Studies' scientific advisor which was used to create the AC536.xlsx file.
    vi. Constellation-bright stars.nb: Mathematica® notebook used to obtain star RA and Declinations which were transformed into cockpit view coordinates.

vii. Final STK Files.zip: is a zip archive of the final STK model used in this paper.
viii. McDowell's RAW TLE 'like' dump.zip: is the zip archive for Jonathan McDowell's Starlink dump which supported early modeling prior to modifications by T.S. Kelso to Celestrak.
ix. STARLINK-8-10.tle: The full list of TLEs for the Group 4-26 Starlink TLEs that were extracted from T.S. Kelso's full TLE supplemental TLE query. These TLEs represent the closest UTC times to the UTC times from both photographs.
x. cockpitview.py: is the Python script used to create our final cockpit view coordinates.
xi. csvout.py: Python script used to process the SOAP extracted data into the comma delimited format.
xii. geoid_height_2023-08-09.csv: Geoid heights from reference (EarthScope Consortium 2023).
xiii. group4-26.zip: is a zip archive of Celestrak's supplemental query results.
xiv. output-analysis.xlsx: is an Excel® spreadsheet with our final analysis results.
xv. relative.py: is a Python script that was initially created to process the csv files, was replaced by cockpitview.py.
xvi. rot.py: is a Python script with the coordinate transformation rotations from Vallado (Vallado 2013).
xvii. sidereal.py: from Michael Hirsch's Python 3-D coordinate conversion code (https://github.com/geospace-code/pymap3d) to support translation of stellar coordinates to the cockpitview coordinate system.
xviii. star_output.csv: the output file from stars.py.
xix. stars.py: my cobbled together python code to create the star_output.csv file which provides stellar coordinates in the cockpit view coordinate system.
xx. vallado.py: from Michael Hirsch's Python 3-D coordinate conversion code (https://github.com/geospace-code/pymap3d) to support translation of stellar coordinates.
xxi. 2024-01435 - FOIA Final Response Letter v2 signed 022924.pdf: FAA's response letter sent to Robert Powell as the result of his FOIA request (Powell 2024).
xxii. Responsive Records for 2024-01435_January 1 2023 to April 30 2023.pdf: FAA's records provided to Robert Powell as the result of his FOIA request (Powell 2024).
xxiii. Responsive Records for 2024-01435_January 1 2023 to April 30 2023.xlsx: Our conversion of the FAA's records into Microsoft Excel™ format provided to Robert Powell as the result of his FOIA request (Powell 2024).
xxiv. FAA FOIA Analysis from Sarah Little.txt: Extracted from private communication.

**In the "SOAP screenshots" subdirectory:**
i. readme.txt: current list of files and the below description of the movie file.
ii. Screenshot 2023-12-25 065808.png: screengrab of a SOAP sim result .

    iii. Screenshot-Photo 1 Cyl Equal Area.png**:** screengrab of a SOAP sim result.
    iv. Screenshot-Photo 1 Geometry.png: screengrab of a SOAP sim result.
    v. Screenshot-Photo 1 Sun2Earth.png: screengrab of a SOAP sim result.
    vi. Screenshot-Photo 2 Cyl Equal Area.png: screengrab of a SOAP sim result.
    vii. Screenshot-Photo 2 Geometry.png: screengrab of a SOAP sim result.
    viii. Screenshot-Photo 2 Sun2Earth.png: screengrab of a SOAP sim result.
    ix. Photo 1 Cyl Equal Area-zoom insert.png: screengrab of a SOAP sim result.
    x. Photo 1 Cyl Equal Area-zoom.png: screengrab of a SOAP sim result.
    xi. Photo 1 Geometry-zoom.png: screengrab of a SOAP sim result.
    xii. Photo 1 UTC time with an added alt error (Fig-6).png: screengrab of a SOAP sim result where 2500 feet (0.8 km) was added to the aircraft's altitude.
    xxv. Photo 2 UTC time with an added alt error.png: screengrab of a SOAP sim result where 2500 feet (0.8 km) was added to the aircraft's altitude.

2. Additional videos and screen shots of our simulations are available on Google Drive at (Buettner, et al. 2024):

   **In the "Movies" subdirectory:**
       i. README.txt: current list of files and the below description of the movie file.
       ii. STK-SatelliteDeployment.mp4: STK simulation of launch through deployment.
       iii. STK-StationarySatelliteFlyBy.mp4: STK simulation of location of Starlink Train w/ respect to a stationary aircraft.
       iv. STK-SatelliteFlyByV2.mp4: STK simulation of the FULL location of Starlink Train w/ respect to the aircraft.
       v. SOAP-AC536-Earth-Sun_movie.wmv: SOAP Earth Polar view showing AC536, G4-26, and the location of the sun-Left, and our standard Earth Ground - Perspective with AC536 in the center with the satellite train G4-26 and the constellations in the background. The atmospheric glow from the sun is displayed as well.
       vi. SOAP-AC536-Earth-Sun_with altitude error movie.wmv: SOAP earth, sun, AC536 perspective with the G4-26 satellite train and ground tracks. The solar terminator is viewed on the edge of the earth's limb and constellations on the left. Our standard Earth Ground - Perspective with AC536 in the center with the satellite train G4-26 and the constellations in the background on the right.
       vii. SOAP-AC536-perspective.wmv: SOAP standard Earth Ground - Perspective with AC536 in the center with the satellite train G4-26 and the

       constellations in the background. The atmospheric glow from the sun is displayed as well.

viii. <u>SOAP-AC536-perspective-movie.wmv</u>: SOAP polar view with the AC536, the ADS-B ground track, and the location of G4-26 on the left, and the standard Earth Ground - Perspective with AC536 in the center with the satellite train G4-26 and the constellations in the background on the right, this time with the AC536 ground track.

**In the "Blender" subdirectory:**

i. <u>Blender Rendering Experiments.zip</u>: is a zip archive of the rendering experiments attempted in this paper (the archive includes all the .blend and .blend1 files with the Boeing CAD model and surrogate 'cubes' for all 52 Starlinks for Blender 4.0.1 and the Earth TIFF file that we used).

ii. <u>Blender Rendering Experiments-images.zip</u>: is a zip archive of our rendering results with other screengrabs of settings and cockpit view location).